# Modeling Ride-Sourcing Matching and Pickup Processes based on Additive Gaussian Process Models

Zheng Zhu[1,2], Meng Xu[3], Yining Di[3], Xiqun (Michael) Chen[1,2*], Jingru Yu[1]

1. College of Civil Engineering and Architecture, Zhejiang University, Hangzhou, China

2. Alibaba-Zhejiang University Joint Research Institute of Frontier Technologies, Hangzhou, China

3. Department of Civil and Environmental Engineering, Hong Kong University of Science and Technology, Hong Kong, China

* Corresponding Author. Email: chenxiqun@zju.edu.cn

**Abstract**: Matching and pickup processes are core features of ride-sourcing services. Previous studies have adopted abundant analytical models to depict the two processes and obtain operational insights; while the goodness of fit between models and data was dismissed. To simultaneously consider the fitness between models and data and analytically tractable formations, we propose a data-driven approach based on the additive Gaussian Process Model (AGPM) for ride-sourcing market modeling. The framework is tested based on real-world data collected in Hangzhou, China. We fit analytical models, machine learning models, and AGPMs, in which the number of matches or pickups are used as outputs and spatial, temporal, demand, and supply covariates are utilized as inputs. The results demonstrate the advantages of AGPMs in recovering the two processes in terms of estimation accuracy. Furthermore, we illustrate the modeling power of AGPM by utilizing the trained model to design and estimate idle vehicle relocation strategies.

## 1. Introduction

The emerging ride-sourcing services provided by transportation network companies (TNCs), such as Uber, Lyft, Didi Chuxing, and Grab, have changed how people travel. Through a few simple clicks and taps on the online platforms (or smartphone apps), people can request ride services at any time and any place. The efficiency in accommodating people's mobility needs makes ride-sourcing a popular mode among travelers. It was reported that Uber had operated services in 24 countries[1], and Didi Chuxing's daily order had exceeded 50 million in August 2020 under the COVID-19 pandemic[2]. Due to the indispensability of ride-sourcing services in modern transportation systems, a solid understanding and suitable modeling of ride-sourcing markets are critical for achieving efficient ride-sourcing services and sustainable urban mobility.

---

[1] Reported on Uber's official website. https://www.uber.com/newsroom/5billion-2/.

[2] News from https://www.yicaiglobal.com/news/didi-chuxing-daily-orders-exceed-50-million-as-china-recovers-

Among a few specific properties, such as the free entry and leave of drivers and accommodating real-time travel demand from passengers, the order matching and passenger pickup processes make ride-sourcing unique compared with other travel modes. Upon receiving a travel request from a passenger, the platform assigns the order to a near idle vehicle, which then picks up and delivers the passenger. Due to the efficient matching via smartphone apps, the meeting friction between passengers and drivers in ride-sourcing markets is much smaller than that in traditional street-hailing taxi markets or dial-hailing taxi markets. Although the platform adopts certain algorithms/rules to match passengers and drivers, the matching and pickup processes are complex and thus difficult to be fully characterized and modeled. From a spatial-temporal perspective, it is complicated to model drivers' and passengers' behaviors due to the imbalance between supply and demand: if the number of idle vehicles near one zone is insufficient to accommodate the travel demand of this zone, the platform needs to call remote drivers to serve these passengers; once the number of drivers is substantially higher than the passenger demand, some idle vehicles might cruise to other zones to get matched. Furthermore, it is challenging to model the matching and pickup processes via conventional first-come-first-served queueing/stochastic models because the platform generally matches passengers and idle vehicles based on their spatial-temporal information rather than in order of arrival.

Although ride-sourcing has attracted extensive attention from researchers, the spatial-temporal modeling of order matching and passenger pickup processes has not been sufficiently investigated, especially in linking actual data with mathematical models. Efforts have been primarily directed toward the theoretical and empirical analyses of passenger demand, driver supply, order dispatching, dynamic ride-sharing, pricing/subsidy, and the economic/social impact on transportation systems (e.g., see Jung et al., 2016; Battifarano and Qian, 2019; Wang and Yang, 2019; Nourinejad and Ramezani, 2020; Zhu et al., 2020; Xu et al., 2020; Ke et al., 2020; Zhu et al., 2021c; Ke et al., 2021). Most system-wide analyses of ride-sourcing markets rely on certain mathematical models (Yang et al., 2010; Cachon et al., 2017), queueing models (Taylor et al., 2017; Zhang et al., 2020), or simulation-based platforms (Holler et al., 2019) to model the matching and pickup processes. Although they obtained analytical/numerical insights on different aspects of the market, the findings could be inaccurate due to the lack of data support and sound frameworks that comprehensively ascertain the distinctive spatial-temporal processes.

This study is among the pioneering works that model spatial-temporal order matching and passenger pickup processes in ride-sourcing markets via methods with data-driven features and analytically tractable formulations. The processes are modeled by the summation of several independent Gaussian Processes (GPs), referred to as the additive Gaussian Process Model (AGPM). A GP is a powerful method for modeling stochastic processes with a target variable as the output and several covariates as the input (Williams, 2006). With high flexibility in building the covariance between variables under the probability theory, AGPM models capture the probabilistic relationship between variables without requiring solid assumptions, which leads to good predictive power and closed-form expressions between variables. Based on a ride-sourcing dataset collected in Hangzhou, China, we develop several AGPMs to fit the matching and pickup processes, in which the number of matches or pickups are used as the output, and spatial, temporal, demand, and supply covariates are utilized as the input. We also train models for comparison based on both widely-used machine-learning approaches and analytical approaches (e.g., queueing models, economic models with closed formulation). The results indicate that AGPMs generally show better goodness of fit for the two processes than machine learning models and analytical models. By incorporating spatial-temporal demand and supply information as exogenous inputs, AGPMs can conduct many market analysis work (e.g., sensitivity analyses and the design of operational strategies) that were previously done by analytical or simulation-based approaches. We utilize the trained AGPM to model idle vehicle relocation strategies and show that closed formulation and analytically tractable derivatives provide insightful guidance for designing strategies that improve the matching efficiency of the market. Thereby, the modeling power of the AGPM is also demonstrated.

The remainder of this paper is organized as follows. **Section 2** reviews the literature on the modeling of matching and pickup processes in ride-sourcing markets. To incorporate widely-used data-driven approaches into our study, we also review the applications of machine learning models and GP models in transportation studies. **Section 3** presents the research objectives of this paper, which is to propose a data-driven approach with both high-level data fitting and modeling power. We also briefly introduce the AGPM and the modeling of ride-sourcing processes via the AGPM. Numerical studies based on a real-world ride-sourcing dataset are conducted in **Section 4**, in which we show both empirical analyses and fitting results of AGPMs (also other machine learning and analytical models) for the spatial-temporal matching and pickup processes. **Section 5** adopts the trained AGPM for modeling and designing idle vehicle relocation strategies. **Section 6** concludes the paper.

## 2. Literature Review

Existing literature commonly employs some analytical approach to demonstrate the process of ride-sourcing market matching and/or pickup, in which different matching functions are adopted based on specific assumptions. Cachon et al. (2017) assumed a random perfect matching rule that ignores the meeting friction, i.e., the number of matched driver-passenger pairs equals the minimum of passenger demand and driver supply. A similar matching model was used by Yu et al. (2020), where they assumed consumer waiting time to be zero. Yang et al. (2010) and Zha et al. (2016) used the Cobb-Douglas-based production function[3] as the passenger/taxi meeting function to derive the bilateral

---

[3] The Cobb–Douglas production function is $Y = AL^{\beta}K^{\alpha}$, where $Y$ denotes the total production, $A$ the productivity factor, $L$ the labor input, $K$ capacity, $\alpha$ and $\beta$ are elasticities. In ride-sourcing studies, the total

passenger-side/driver-side waiting time as well as the number of order matches. Banerjee et al. (2016) adopted an M/M1 queueing model to estimate the idle time of ride-sourcing vehicles. Taylor et al. (2017) and Bai et al. (2018) used M/M/K queueing model to approximate the customer's expected waiting time. Xu et al. (2020) built a double-ended queuing model to simulate the behavior of the platform by introducing request sequences within a matching block. Zhang et al. (2019) developed a spatial queueing network and estimated the Poisson rate of demand and supply via New York taxi data, while queueing models are somehow incapable of modeling cases where passenger demand is much greater than the service capacity. Castillo et al. (2017) used a model under the guidance of the first-dispatch protocol, which assigns drivers to respond to passengers' requests for travel immediately. Feng et al. (2021) proposed mathematic functions to model driver and passenger matches on a circle network. Concerning the many-to-many situation in the matching pool, Xu et al. (2017) proposed a dynamic batch matching model. Zhu et al. (2020) combined the Cobb-Douglas-based production function with a detour function and estimated passenger waiting time and detour time for a market with both ride-sourcing and ride-sharing. Moreover, the matching mechanisms have different objectives, such as minimizing costs (Baldacci et al., 2004) and maximizing the number of served passengers (Agatz et al., 2011; Masoud and Jayakrishnan, 2017). Readers can refer to Tafreshian et al. (2020) for more details about ride-sourcing and ride-sharing matching mechanisms.

Without any doubt, these studies contribute to operational and management analyses of ride-sourcing markets; yet a common issue is that most of the analytical models have not been validated by real-world data. Thereby, it is not clear whether the aforementioned models are appropriate in depicting ride-sourcing matching and pickup processes.

To model the actual matching and pickup processes in ride-sourcing markets, a general idea is to develop regression/data-driven/machine learning models based on real-world data. We summarize several widely-used machine learning methods in transportation research that have shown good performance.

- Decision tree (DT) models and extensions, e.g., random forest (RF) models and boosted DTs. DT models and the family have been used in different transportation studies and achieved good results, such as the detection of users' travel modes (Zhou et al., 2016), short-term prediction of ride-sourcing demand (Saddi et al., 2017), estimating the probability of choosing ride-sharing among different ride-sourcing service options (Zhu et al., 2021b). RF and boosted DT models tend to provide better fitting results than vanilla DTs, but they sacrifice interpretability to a certain degree via obtaining groups of DTs from the data and bagging them with random subsets.

- Support vector machine (SVM) models. In transportation-related machine learning studies, Zhang et al. (2007), Yao et al. (2012), and Ganji et al. (2020) used SVM for freeway traffic volume prediction; Wu et al. (2004) adopted SVM to predict

---

production is replaced by the number of matches, labor, and capacity are replaced by passenger demand and driver supply.

highway estimated time of arrival; Tseng et al. (2018) built a real-time highway congestion prediction model based on SVM which utilized the streamflow data such as section speed, road density, and road traffic volume; Allahviranloo et al. (2013) used multiclass-SVM to learn the parameters of daily activity engagement patterns. SVM is powerful in addressing local minima and overfitting problems, but it cannot model multiple classification problems and has poor interpretability.

- Neural network (NN) models and extensions, e.g., recurrent neural network (RNN), convolutional neural network (CNN), and graph neural network (GNN), have gradually received attention and been utilized in various transportation areas (Karlaftis and Vlahogianni, 2011). Cantarella and Luca (2005) used multi-layer NN to infer travel mode choices. Zeng and Zhang (2013) proposed an RNN-based model to predict freeway travel time. Yao et al. (2018) utilized a CNN-embedded RNN for spatial-temporal taxi demand prediction. Wang et al. (2020) and Peng et al. (2020) adopted GNN and dynamic GNN, respectively, for traffic flow prediction. It should be noted that, compared to conventional NNs, deep learning methods, e.g., RNN and CNN, address the problem of long training time and overfitting to some extent, but they still face the issue of poor interpretability.

The common issue of machine learning methods lies in the weakness of interpreting the quantitative relationship between the target variable (output) and the related covariates (input). This can be improved by using statistical methods such as the GP model. Recently, the application of GP models to spatial-temporal prediction tasks has drawn heated attention. Kupilik and Witmer (2018) adopted a conventional GP regression model to predict violence in Africa. Venkitaraman et al. (2020) proposed a graph-based GP model and applied it for temperature prediction in 45 cities in Sweden. Based on additive kernels with spatial and temporal features, Senanayake et al. (2016) used GP models to predict the trend of flu in the US; Cheng et al. (2019) applied additive GP models to regressing several longitudinal datasets; Ma et al. (2019) proposed a large-scale GP for air quality prediction; Sheibani and Ou (2021) utilized GP for damage inference under earthquakes. In transportation studies, GP-based methods have been applied in travel time prediction (Id'e and Kato, 2009), public transport demand prediction (Neumann et al., 2009), traffic volume prediction (Xie et al., 2010), traffic speed imputation (Rodrigues and Pereira, 2018; Yuan et al., 2021). However, like other machine learning methods, GPs have not been applied to modeling ride-sourcing order matching and passenger pickup processes.

In summary, in modeling ride-souring matching and pickup processes: 1) most analytical approaches have not been linked with actual ride-sourcing data; 2) many widely-used machine learning models and GP models have not been applied in fitting/modeling these processes. This paper attempts to fill the research gap by fitting these ride-sourcing processes via AGPMs, analytical models, and machine learning models based on real-world data. Moreover, we show the capability of the AGPM in designing and analyzing idle vehicle relocation strategies, which demonstrates that the AGPM is a promising approach for data-driven ride-sourcing market modeling.

## 3. Modeling Ride-Sourcing Markets via the Additive Gaussian Process Model

This section first provides the problem statement and objective of data-driven ride-sourcing market modeling. Then, we introduce the conventional GP regression model (or GP for short) and the AGPM about the formulation in fitting and predicting spatial-temporal stochastic processes. Finally, we conceptually show the spatial-temporal modeling of ride-sourcing markets via the AGPM.

### 3.1. Problem Statement for Data-Driven Ride-Sourcing Market Modeling

A ride-sourcing market, which involves spatial-temporal interactions between the platform, drivers, passengers, and other transportation features, is complicated to characterize and model. As mentioned in **Section 2**, researchers have proposed analytical approaches with closed formulation, such as queueing models and economic models, to depict the relationship between different exogenous variables; and based on these models, they have analyzed key attributes of the market and gained operational and management insights. However, insufficient data support has limited these models' capability to describe unique processes accurately (e.g., spatial-temporal order matching, passenger pick, and idly cruising) in ride-sourcing markets.

We aim to propose a general approach to modeling spatial-temporal ride-sourcing processes based on actual data. The method for developing the data-driven ride-sourcing model shall be powerful in fitting/prediction accuracy and have analytically tractable expressions between endogenous variables and exogenous variables, as well as the gradient (derivative) information of exogenous variables. Closed-form mathematical expressions and derivatives are necessary for gaining analytical and management insights in market analysis, marginal effect analysis, and system operation tasks, such as designing and optimizing operational strategies. The mathematical description of our research problem is provided below.

Let $y$ denote the endogenous (output) variable in the ride-sourcing market, such as the number of matches, passenger waiting time, or the number of pickups; $Y = [y_1, y_2, \dots, y_N]^\mathsf{T}$ denotes the collection of endogenous variables, where $N$ is the sample size. Let row vector $x = [x^1, x^2, \dots, x^R] \in \chi$ denote spatial-temporal exogenous (input) covariates, such as the number of passengers, number of drivers, traffic condition, or weather information, where $R$ is the input dimension, and $\chi$ is the input domain; $X = [x_1^\mathsf{T}, x_2^\mathsf{T}, \dots, x_N^\mathsf{T}]^\mathsf{T}$ represents the collection of inputs corresponding to $Y$. Note that we use bold font for vectors/matrices and non-bold font for scalers/functions in this paper. The objectives are to:

- Fit an analytically tractable function $y = g(x)$ such that the gap between predicted outputs and observed outputs is minimized, i.e., $\min(\sum_{i=1}^{N} \|y_i - g(x_i)\|)$.
- Obtain closed-form derivatives $\frac{\mathrm{d}y}{\mathrm{d}x^r} = \frac{\mathrm{d}g(x)}{\mathrm{d}x^r}$, $r \in \{1, 2, \dots, R\}$ for marginal effect analysis and operations research tasks.

Many machine learning models could be unsuitable for our research problem since they are black boxes in depicting mathematical relationships between variables. Besides, the previously developed analytical ride-sourcing models could be weak in the goodness of fit. AGPM, a flexible and non-parametric statistical approach, is well desirable for achieving our objectives. Furthermore, since the attributes of ride-sourcing markets (e.g., passenger demand, driver supply, price, and subsidies) are of high stochasticity in spatial and temporal dimensions, AGPM could be an appropriate alternative to modeling stochastic processes in ride-sourcing markets.

### 3.2. The Additive Gaussian Process Model

A Gaussian process (GP) is the probability distribution of nonlinear functions. Let $f(\boldsymbol{x})$ be a function, and a GP is defined as

$$f(\boldsymbol{x}) \sim \text{GP}\big(m(\boldsymbol{x}), k(\boldsymbol{x}, \boldsymbol{x}')\big), \forall \boldsymbol{x}, \boldsymbol{x}' \in \chi \tag{1}$$

where $m(\boldsymbol{x}) = \text{E}\big(f(\boldsymbol{x})\big)$ denotes the mean function value at input vector $\boldsymbol{x}$; and $k(\boldsymbol{x}, \boldsymbol{x}') = \text{Cov}\big(f(\boldsymbol{x}), f(\boldsymbol{x}')\big)$, which is a positive-semidefinite function, represents the covariance between any two realizations of $f(\boldsymbol{x})$ and $f(\boldsymbol{x}')$. The mean $m(\boldsymbol{x})$ is usually set as zero, and $k(\boldsymbol{x}, \boldsymbol{x}')$ is referred to as a kernel function (or kernel for short).

Function values of input matrix $\boldsymbol{X}$ is denoted by $f(\boldsymbol{X}) = [f(\boldsymbol{x}_1), f(\boldsymbol{x}_2), \dots, f(\boldsymbol{x}_N)]^\intercal$ and they follow a joint multi-variate Gaussian distribution:

$$f(\boldsymbol{X}) \sim \text{N}\left(\boldsymbol{0}, \boldsymbol{K}_{\boldsymbol{X}, \boldsymbol{X}}(\boldsymbol{\theta})\right) \tag{2}$$

where $\text{N}(\boldsymbol{\mu}, \boldsymbol{\Sigma})$ denotes the multi-variate Gaussian distribution with mean $\boldsymbol{\mu}$ and covariance matrix $\boldsymbol{\Sigma}$, $\boldsymbol{K}_{\boldsymbol{X}, \boldsymbol{X}}(\boldsymbol{\theta})$ denotes the N-by-N covariance matrix such that $\big[\boldsymbol{K}_{\boldsymbol{X}, \boldsymbol{X}}(\boldsymbol{\theta})\big]_{i,j} = k\big(\boldsymbol{x}_i, \boldsymbol{x}_j | \boldsymbol{\theta}\big)$, and $\boldsymbol{\theta}$ represents the parameter vector of the kernel.

GPs can be utilized to depict the probabilistic relationship between $\boldsymbol{x}$ and $y$, which is given by the following hierarchical model:

$$\boldsymbol{\theta} \sim \pi(\boldsymbol{\phi}) \tag{3}$$

$$f(\boldsymbol{X}) \sim \text{N}\left(\boldsymbol{0}, \boldsymbol{K}_{\boldsymbol{X}, \boldsymbol{X}}(\boldsymbol{\theta})\right) \tag{4}$$

$$Y \sim \text{N}(f(\boldsymbol{X}), \sigma_\varepsilon^2 \boldsymbol{I}_N) \tag{5}$$

where $\sigma_\varepsilon^2$ is the noise variance that is included in parameter vector $\boldsymbol{\theta}$, $\pi(\boldsymbol{\phi})$ represents the prior information of the parameter vector, and $\boldsymbol{I}_N$ denotes an N-by-N identity matrix. Note that the model is equivalent to $y = f(\boldsymbol{x}) + \varepsilon$, where $\varepsilon$ is the noise with variance $\sigma_\varepsilon^2$. We can marginalize $f(\boldsymbol{X})$ analytically and obtain the following model:

$$Y \sim \text{N}\big(\boldsymbol{0}, \boldsymbol{K}_{\boldsymbol{X}, \boldsymbol{X}}(\boldsymbol{\theta}) + \sigma_\varepsilon^2 \boldsymbol{I}_N\big) \tag{6}$$

By learning the parameters, one can use the trained GP for various purposes such as estimation, prediction, missing value imputation, and analyses. Given the learned parameter $\boldsymbol{\theta}$ and a new collection of input vectors $\boldsymbol{X}^* = \left[\boldsymbol{x}_1^{*\top}, \boldsymbol{x}_2^{*\top}, \dots, \boldsymbol{x}_M^{*\top}\right]^\top$, the predicted output vector $\widehat{\boldsymbol{Y}}^* = [\hat{y}_1^*, \hat{y}_2^*, \dots, \hat{y}_M^*]^\top$ has the following joint distribution:

$$\widehat{\boldsymbol{Y}}^* \sim N(\boldsymbol{\mu_\theta}, \boldsymbol{\Sigma_\theta}) \tag{7}$$

$$\boldsymbol{\mu_\theta} = \boldsymbol{K}_{\boldsymbol{X}^*,\boldsymbol{X}}(\boldsymbol{\theta})\big(\boldsymbol{K}_{\boldsymbol{X},\boldsymbol{X}}(\boldsymbol{\theta}) + \sigma_\varepsilon^2 \boldsymbol{I}_N\big)^{-1}\boldsymbol{Y} \tag{8}$$

$$\boldsymbol{\Sigma_\theta} = \boldsymbol{K}_{\boldsymbol{X}^*,\boldsymbol{X}^*}(\boldsymbol{\theta}) - \boldsymbol{K}_{\boldsymbol{X}^*,\boldsymbol{X}}(\boldsymbol{\theta})\big(\boldsymbol{K}_{\boldsymbol{X},\boldsymbol{X}}(\boldsymbol{\theta}) + \sigma_\varepsilon^2 \boldsymbol{I}_N\big)^{-1}\boldsymbol{K}_{\boldsymbol{X},\boldsymbol{X}^*}(\boldsymbol{\theta}) \tag{9}$$

where $\boldsymbol{K}_{\boldsymbol{X}^*,\boldsymbol{X}}(\boldsymbol{\theta})$ is an M-by-N matrix such that $\left[\boldsymbol{K}_{\boldsymbol{X}^*,\boldsymbol{X}}(\boldsymbol{\theta})\right]_{i,j} = k(\boldsymbol{x}_i^*, \boldsymbol{x}_j|\boldsymbol{\theta})$ and $\boldsymbol{K}_{\boldsymbol{X},\boldsymbol{X}^*}(\boldsymbol{\theta}) = \boldsymbol{K}_{\boldsymbol{X}^*,\boldsymbol{X}}(\boldsymbol{\theta})^\top$, $\boldsymbol{K}_{\boldsymbol{X}^*,\boldsymbol{X}^*}(\boldsymbol{\theta})$ is an M-by-M matrix with $\left[\boldsymbol{K}_{\boldsymbol{X}^*,\boldsymbol{X}^*}(\boldsymbol{\theta})\right]_{i,j} = k(\boldsymbol{x}_i^*, \boldsymbol{x}_j^*|\boldsymbol{\theta})$.

To increase the flexibility of the covariance structure and define a more interpretable model, we consider an additive Gaussian Process Model (AGPM) with $D$ independent components:

$$f(\boldsymbol{x}) = f^{(1)}(\boldsymbol{x}) + f^{(2)}(\boldsymbol{x}) + \cdots + f^{(D)}(\boldsymbol{x}) \tag{10}$$

$$y = f(\boldsymbol{x}) + \varepsilon \tag{11}$$

where each $f^{(d)}(\boldsymbol{x}) \sim \text{GP}\left(0, k^{(d)}(\boldsymbol{x}, \boldsymbol{x}')\right)$ is a separate GP with a specific kernel and parameters $\boldsymbol{\theta}^{(d)}$ and $\varepsilon$ is the noise with variance $\sigma_\varepsilon^2$.

In many regression or data-driven studies (e.g., see Battifarano and Qian, 2019; Zhu et al., 2021b), researchers aim to analyze multi-dimensional input that contains spatial information, temporal information, as well as other covariates. To identify the contribution of different covariates to the summation GP $f(\boldsymbol{x})$, we assume that each GP $f^{(d)}(\boldsymbol{x})$ depends on only a few covariates. That is, $f^{(d)}(\boldsymbol{x}) = f^{(d)}\big(\boldsymbol{x}^{(d)}\big) \sim \text{GP}\left(0, k^{(d)}\big(\boldsymbol{x}^{(d)}, \boldsymbol{x}^{(d)'}\big)\right)$, $\boldsymbol{x}^{(d)} \in \chi^{(d)} = \prod \chi^j$, $j \in J_d \subseteq \{1, 2, \dots, R\}$, where $\chi^j$ denotes the domain of the $j$th covariate and $J_d$ represents indices of the covariates associated with the $d$th GP and kernel.

Since the summation of multi-variate Gaussian random variables is still a Gaussian, we know that $f(\boldsymbol{x})$ follows multi-variate Gaussian distribution. Let $\boldsymbol{\theta} = \left[\boldsymbol{\theta}^{(1)}, \boldsymbol{\theta}^{(2)}, \dots, \boldsymbol{\theta}^{(D)}, \sigma_\varepsilon^2\right]$ denote the parameter vector of AGPM, we obtain the marginal likelihood for the output collection $\boldsymbol{Y} = [y_1, y_2, \dots, y_N]^\top$ by marginalizing $f(\boldsymbol{x})$, given by

$$\text{p}(Y|X,\Theta) = \text{p}_N\left(\boldsymbol{Y}\middle|\boldsymbol{0}, \sum_{d=1}^{D} \boldsymbol{K}_{\boldsymbol{X},\boldsymbol{X}}^{(d)}\big(\boldsymbol{\theta}^{(d)}\big) + \sigma_\varepsilon^2 \boldsymbol{I}_N\right) \tag{12}$$

where $\text{p}(\cdot)$ is the probability, $\text{p}_N(\boldsymbol{Y}|\boldsymbol{\mu}, \boldsymbol{\Sigma})$ denotes the probability density value of multi-variate Gaussian distribution at $\boldsymbol{Y}$ with mean $\boldsymbol{\mu}$ and covariance matrix $\boldsymbol{\Sigma}$, and $\boldsymbol{K}_{\boldsymbol{X},\boldsymbol{X}}^F(\boldsymbol{\theta}) = \sum_{d=1}^{D} \boldsymbol{K}_{\boldsymbol{X},\boldsymbol{X}}^{(d)}\big(\boldsymbol{\theta}^{(d)}\big)$ can be referred to as the "full kernel".

Given observed data, we aim to estimate $\boldsymbol{\theta}$ by maximizing the marginal likelihood, which can be done by numerical algorithms such as "BFGS" and "L-BFGS-B".

Alternatively, we can assume $\boldsymbol{\theta}$ to be random variables that follow some prior probability distributions (like in Eq. (3)), and use Bayesian methods to estimate the posterior distribution of $\boldsymbol{\theta}$. The Bayesian approach could produce more robust results that overcome overfitting issues. With estimated parameters, similar to a conventional GP, one can make predictions on newly observed $\boldsymbol{X}^*$ by replacing $\boldsymbol{K}_{X,X}(\boldsymbol{\theta})$, $\boldsymbol{K}_{X,X^*}(\boldsymbol{\theta})$, $\boldsymbol{K}_{X^*,X}(\boldsymbol{\theta})$, and $\boldsymbol{K}_{X^*,X^*}(\boldsymbol{\theta})$ in Eqs. (7–9) with the matrices calculated via "full kernels" $\boldsymbol{K}_{X,X}^F(\boldsymbol{\theta})$, $\boldsymbol{K}_{X,X^*}^F(\boldsymbol{\theta})$, $\boldsymbol{K}_{X^*,X}^F(\boldsymbol{\theta})$, and $\boldsymbol{K}_{X^*,X^*}^F(\boldsymbol{\theta})$ (Eq. (12)).

Many kernel function types can be implemented to build an AGPM with covariates related to temporal and spatial information. We list several kernel functions that reflect common domain knowledge of spatial-temporal study designs in **Appendix A**, and some of them are utilized and compared in developing the ride-sourcing order matching and passenger pickup AGPMs in the case study of this paper.

### 3.3. Modeling Ride-Sourcing Market via AGPMs

We can build a ride-sourcing AGPM by treating spatial-temporal demand and supply as exogenous (input) vector $\boldsymbol{x} = [x^1, x^2, \dots, x^R]$ and attributes of the ride-sourcing processes (e.g., the number of matches/pickups) as the endogenous (output) variable $y$. Based on real-world ride-sourcing data, we train the AGPM by maximizing the marginal likelihood with parameter set $\boldsymbol{\theta}$ to obtain an analytically tractable function $y = g(\boldsymbol{x})$.

The trained AGPM is far more than a non-parametric regression tool; instead, it is a generalization of the stochastic ride-sourcing process. In addition to demand and supply prediction tasks, AGPM can be used in the same way as many previously developed models for analysis, operation, and simulation research tasks (e.g., Zha et al., 2016; Chen and Nie, 2017; Ramezani and Nourinejad, 2018; Wei et al., 2020). Similar to an analysis model of the ride-sourcing market, we assume the relationship between supply and demand variables follows the probabilistic relationship characterized by the AGPM, i.e., $y = g(\boldsymbol{x})$. By treating input variables as decision variables, we can design operational strategies and estimate the potential market performance. Moreover, the analytically tractable model allows us to obtain the marginal effects of certain input variables on the output by taking derivatives $\frac{\mathrm{d}y}{\mathrm{d}x^r}$, $r \in \{1, 2, \dots, R\}$, which is infeasible for other machine learning models.

We provide a naïve example to illustrate how such a stochastic model can be used for modeling/analyzing ride-sourcing markets. In the model developed by Zha et al. (2016), both the number of passengers and the number of ride-sourcing vehicles are exogenous variables; and the Cobb-Douglas-based production function is used to estimate the endogenous number of matches and passenger-driver matching time. Similarly to this example, we could develop and train an AGPM with one squared exponential (SE) kernel, in which $\boldsymbol{x} = [x^d, x^s]$ ($x^d$ and $x^s$ denote numbers of passengers and vehicles, respectively) is the exogenous vector and $y^m$ (the number of order matches) is the endogenous variable. The trained AGPM provides a closed-form stochastic function denoted as $y^m = g(\boldsymbol{x})$, such that the mean and variance of $y^m$ at specific input $\boldsymbol{x}^* = [x^{d*}, x^{s*}]$ are given by $\mathrm{E}(g(\boldsymbol{x})) = \boldsymbol{K}_{X^*,X}^F(\boldsymbol{\theta})\boldsymbol{R}_Y^F(\boldsymbol{\theta})\boldsymbol{Y}$ and $\mathrm{Cov}(g(\boldsymbol{x})) = \boldsymbol{K}_{X^*,X^*}(\boldsymbol{\theta}) - \boldsymbol{K}_{X^*,X}^F(\boldsymbol{\theta})\boldsymbol{R}_Y^F(\boldsymbol{\theta})\boldsymbol{K}_{X,X^*}^F(\boldsymbol{\theta})$,

respectively, where $\boldsymbol{R}_Y^F(\boldsymbol{\theta}) = \left(\boldsymbol{K}_{X,X}^F(\boldsymbol{\theta}) + \sigma_\varepsilon^2 \boldsymbol{I}_N\right)^{-1}$ is a constant matrix. Furthermore, we obtain closed-form derivatives with $\boldsymbol{x}^*$, for example, $\frac{\mathrm{dE}(g(\boldsymbol{x}))}{\mathrm{d}x^{d*}} = \boldsymbol{W}_{\boldsymbol{x}^*,\boldsymbol{X}}^d(\boldsymbol{\theta})\boldsymbol{R}_Y^F(\boldsymbol{\theta})\boldsymbol{Y}$, where $\boldsymbol{R}_Y^F(\boldsymbol{\theta})\boldsymbol{Y}$ is an N-by-1 constant column vector and $\boldsymbol{W}_{\boldsymbol{x}^*,\boldsymbol{X}}^d(\boldsymbol{\theta})$ is a 1-by-N row vector with $\left[\boldsymbol{W}_{\boldsymbol{x}^*,\boldsymbol{X}}^d(\boldsymbol{\theta})\right]_i = \frac{x_i^d - x^{d*}}{l_{SE}^2} k_{SE}(\boldsymbol{x}^*, \boldsymbol{x}_i|\boldsymbol{\theta})$ (refer to **Appendix A** for the SE kernel and components $k_{SE}(\cdot)$ and $l_{SE}^2$). Based on the stochastic function and the derivatives, we can analyze the market attributes, such as order matching efficiency under certain exogenous variables; we can also propose fleet size regulation strategies to optimize the number of matches.

In the light but not limited to the naïve example, we can incorporate additional spatial and temporal features into AGPM and develop more comprehensive ride-sourcing models. **Figure 1** provides a flowchart of modeling ride-sourcing processes via AGPMs. First, we determine the research scope and modeling process with sufficient ride-sourcing features, traffic conditions, and other data (e.g., urban population, and weather). Second, we include variables of interest to capture inherent characteristics and complicated relations in the modeling process. Third, AGPMs with different structures and kernels are trained, wherein the one with the best goodness of fit is selected as our final model. Last, the selected trained AGPM stochastic function can be used for operation, analysis, prediction, and simulation research tasks of the ride-sourcing market.

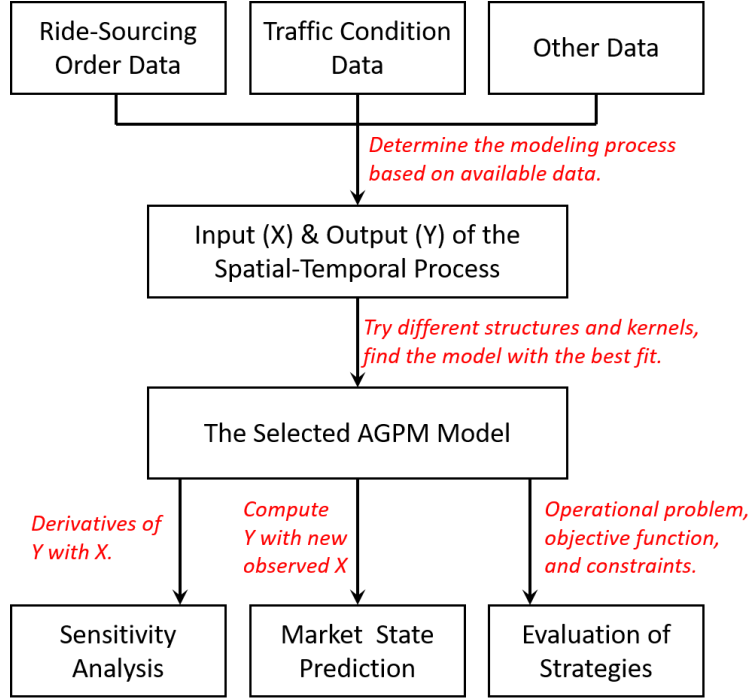

**Figure 1.** A general flowchart of modeling ride-sourcing processes via AGPMs

The AGPM based modeling approach can be superior in the following aspects compared with other ride-sourcing models:

- As a data-driven approach, AGPM has better goodness of fit and provides more accurate estimations than analytical models (e.g., queueing models, economic models).
- Being a probabilistic model, AGPM has closed-form expressions for both the expected and covariance of output variables, as well as their derivatives, which are not available for many machine learning models.
- Since $R_Y^F$ and $R_Y^F Y$ are constants, AGPM is more computationally efficient than simulation-based approaches in large-scale research problems.

To this end, we have proposed a data-driven approach to modeling spatial-temporal processes in ride-sourcing markets. Theatrically, the AGPM-based ride-sourcing modeling approach can accurately reconstruct order matching and passenger pickup processes and provide analytically tractable functions. Empirical performances of AGPM with goodness of fit and modeling capability are given in **Section 4** and **Section 5**.

## 4 Fitting Ride-Sourcing Processes via Real-World Data

This section trains AGPMs for order matching and passenger pickup processes based on a real-world ride-sourcing dataset. The dataset is introduced in **Section 4.1**. We empirically explore the spatial-temporal attributes of the two processes in **Section 4.2**. In **Section 4.3**, AGPMs and other machine learning and analytical approaches are utilized to fit the processes.

### 4.1 Data Description and Processing

The ride-sourcing dataset used in this paper was collected by the Bureau of Transportation in Hangzhou, China. The dataset contains ride-sourcing order records in Hangzhou City from December 3rd to 7th (Monday to Friday), 2018, which are reported by different TNCs (e.g., DiDi, Shenzhou, Shouyue). For each ride-sourcing order, the dataset covers detailed information, including order creating time, scheduled departure time, order matched time, idle vehicle's dispatched location (i.e., the vehicle's latitude/longitude at the order matched time), pickup time, latitude/longitude of the passenger's origin, latitude/longitude of the passenger's destination, passenger delivery time, trip fare and so on. The precision of time is "second", such that one can compute the following measurements for each order:

- Matching time: the duration from order creation time to order match time.
- Pickup time: the duration from order match time to passenger pickup time.
- In-vehicle delivery time: the duration from pickup time to order finish time.

Of particular interest is to recover actual order matching and passenger pickup processes based on this dataset. In a ride-sourcing market, passengers create orders (i.e., request trip services) at different time points and locations and drivers who finish delivering a passenger (i.e., finish an order) become idle at the previous passenger's destination and can be dispatched to serve the next passenger. Therefore, order matching and passenger pickup processes can be regarded as spatial-temporal interactions between passengers and drivers.

**Table 1.** Notations of variables

| Variable | Meaning |
|---|---|
| *Input (exogenous)* | |
| $x^r$ | Row index of the zone (starting from the north) |
| $x^c$ | Column index of the zone (starting from the west) |
| $x^t$ | Time index for small time intervals |
| $x^d$ | Passenger demand in the zone |
| $x^s$ | Number of idle vehicles in the zone |
| *Output (endogenous)* | |
| $y^m$ | Number of matches in the zone |
| $y^p$ | Number of pickups in the zone |

Concerning planning and operational purposes, we utilize small hexagon zones (the radius is 1 km) and short time intervals (3 minutes) as the spatial and temporal analysis unit, respectively, for modeling the two processes. **Table 1** summarizes the spatial, temporal, demand, and supply variables used in our study. We first partition a study area into a network of adjacent hexagon zones; $x^r$ and $x^c$, respectively, denote the row index (starting from the north) and column index (starting from the west) of a hexagon zone in the network. We use $x^t$ to denote the index of the short time intervals in the analysis horizon. By summarizing detailed order records according to the slots of hexagon zones and time intervals, we can obtain the following measurements at each zone and time interval: 1) the number of passenger demand $x^d$ based on order creating time and origin; 2) the number of driver supply $x^s$ based on order finish time and destination (i.e., the finish of an order is counted as an available supply); 3) the number of matched orders $y^m$ according to order match time and origin; 4) the number of pickups $y^p$ by order pickup time and origin. Therefore, the objective is to model $y^m$ and $y^p$ for each zone and time interval using vector $\boldsymbol{x} = [x^r, x^c, x^t, x^d, x^s]$, recovering the spatial-temporal process across the entire study area and time horizon. However, we need to note that order cancelations, which significantly influence the matching and pickup processes (Wang et al., 2020; Xu et al., 2022) are dismissed in this paper because the Hangzhou dataset has not included cancelation information; otherwise, we shall also add the spatial-temporal cancelation rate as an input variable.

### 4.2 Empirical Analyses of Matching and Pickup Processes

This case study focuses on analyses of DiDi's ride-sourcing market in downtown Hangzhou, which is presented as 36 small hexagon zones. **Figure 2** shows the study area and spatial variation of passenger demand and driver supply from 7:00 to 10:00 on December 3rd, 2018. A "zone ID" is assigned to each hexagon unit with the format of "($x^r$, $x^c$)". We use heat maps to present the demand/supply rate (number per minute, i.e., the number of passengers/idle vehicles in the zone during the 3-hour window divided by 180 minutes). From the figure, we observed some supply-demand imbalance. For instance, the demand in zone *(1, 2)* is notably higher than the supply there, while the supply in zone *(4, 3)* is higher than its passenger demand.

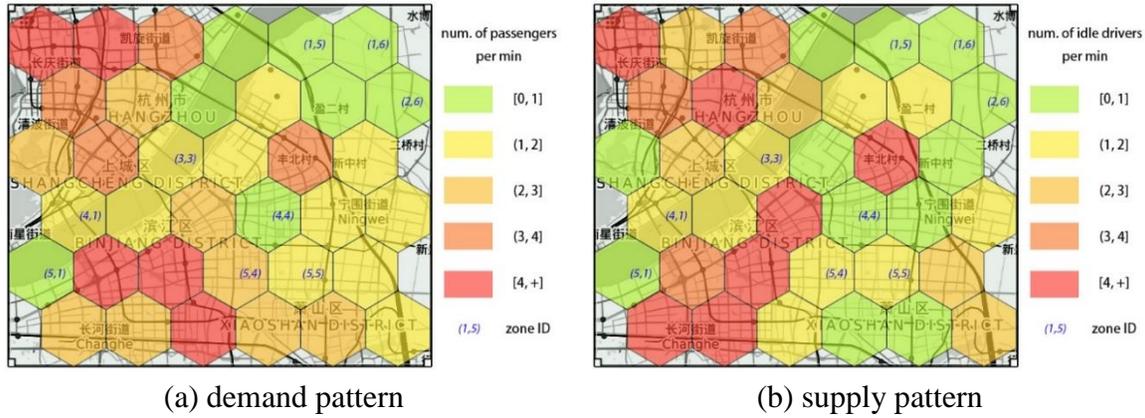

(a) demand pattern           (b) supply pattern

**Figure 2.** Spatial-temporal pattern of demand and supply in downtown Hangzhou

The imbalance between passenger demand and driver supply essentially affects the matching and pickup processes. From the passenger perspective, insufficient supply at one zone could cause a long matching queue at that zone. However, this queueing phenomenon, which is a general setting in many ride-sourcing queueing systems (e.g., Taylor et al. 2017; Xu et al., 2020), cannot accurately describe the market for two reasons. First, ride-sourcing is not a first-in-first-out (FIFO) system, and even if a passenger places an order later than many others, he/she could get a match before them if the next idle vehicle in the system is very close to him/her but far from others. Second, passengers in one zone can be matched with drivers in the same zone and drivers in adjacent or remote zones; once there are sufficient idle vehicles in adjacent zones, passengers could enjoy a short matching time. For the second reason, passengers could suffer from some extra pickup time due to a long pickup distance; and it provides the foundation for modeling approaches that depict the number of passenger-driver meets as functions of demand and supply (Yang et al., 2010; Zha et al., 2016). The length of a pickup queue (i.e., the number of passengers who have gotten a match and are waiting for pickup) is also significantly affected by the traffic condition. Even though there is sufficient supply in congested areas, the pickup queue (also pickup time) could be large.

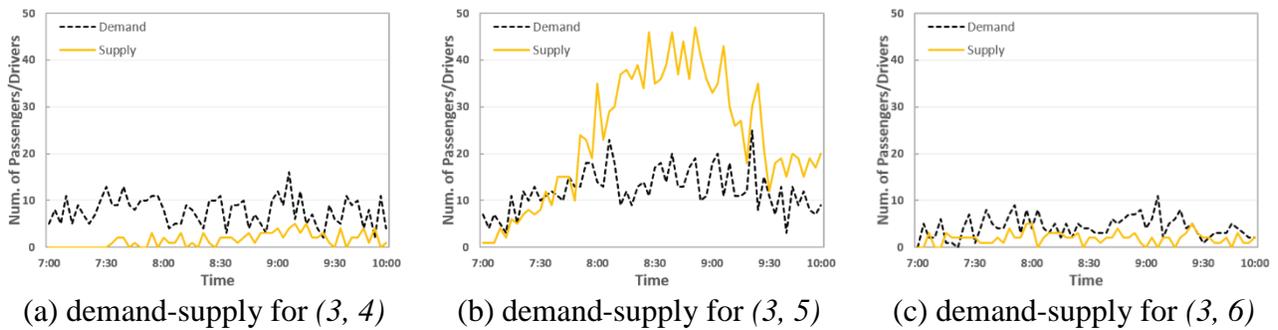

(a) demand-supply for *(3, 4)*     (b) demand-supply for *(3, 5)*     (c) demand-supply for *(3, 6)*

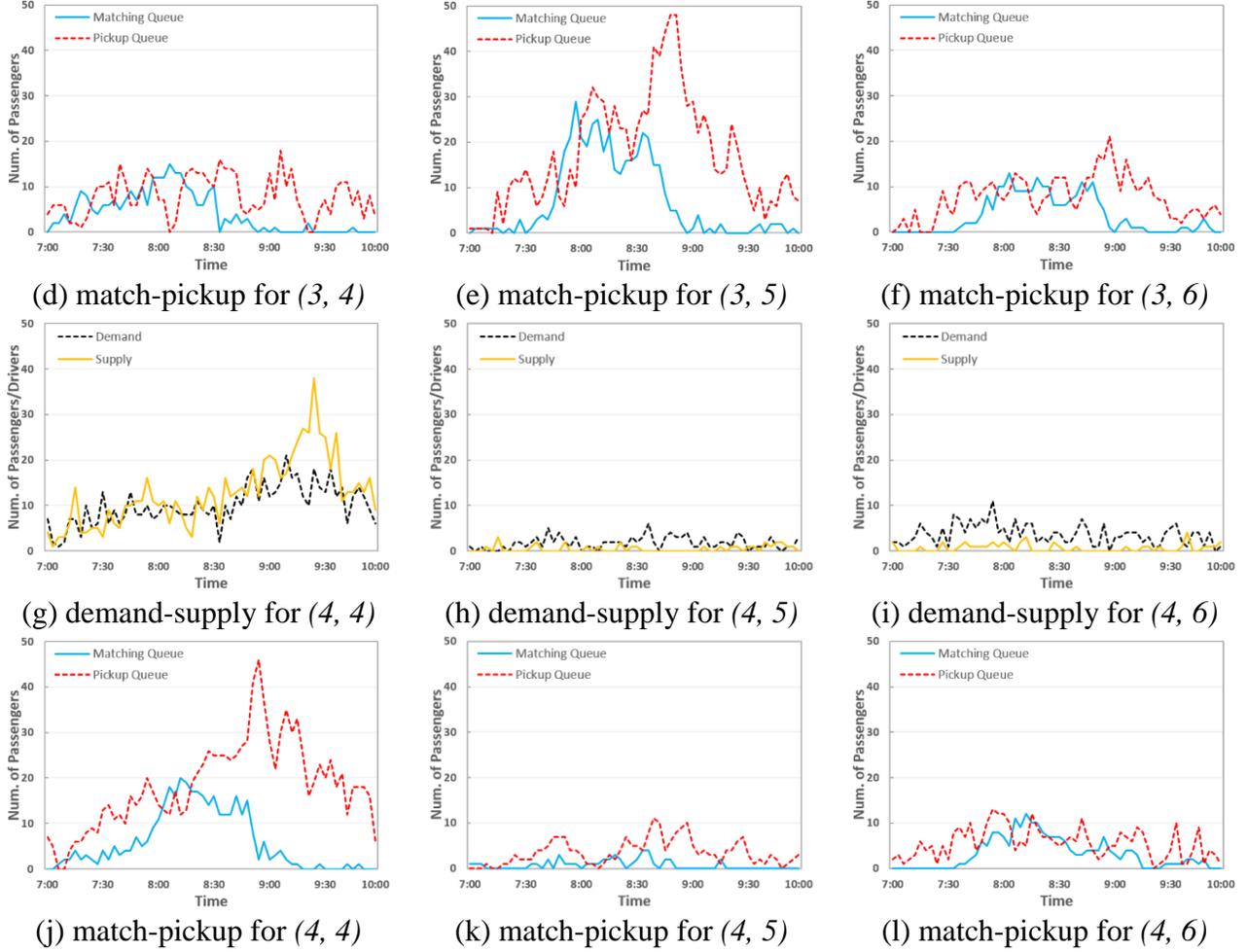

**Figure 3.** Profiles of demand, supply, matching queue, and pickup queue

**Figure 3** illustrates the actual "queueing" process for matching and pickup with the profiles of demand and supply for neighboring zones *(3, 4)*, *(3, 5)*, *(3, 6)*, *(4, 4)*, *(4, 5)*, *(4, 6)* on December 3rd. First, we note that although the demand in zones *(3, 4)*, *(3, 6)*, *(4, 5)*, and *(4, 6)* is notable higher than the supply during 8:30 to 9:30, the matching queues for these zones are short. This is because the supply in the zone *(3, 5)* is abundant and the surplus idle vehicles are dispatched to other adjacent zones for relieving the matching queues. Second, even though the supply in zones *(3, 5)* and *(4, 4)* is much sufficient for matching the demand, passengers in the two zones always encounter severe queueing for pickup. The reason is that the traffic is seriously congested in the two zones, which absorbs vast numbers of passengers during the morning peak period.

In summary, the order matching and passenger pickup processes in ride-sourcing markets are complicated with spatial and temporal interactions. It is imperfect to model the two processes as general queueing models or aggregate models. A sound model is required to incorporate spatial-temporal heterogeneity of the market (network) and the relationship among key measurements, e.g., demand, supply, the number of matches, the number of pickups, etc.

### *4.3 Fitting Matching and Pickup Processes*

To qualitatively and quantitatively understand the ride-sourcing matching and pickup processes, we treat $y^m$ or $y^p$ as outputs and $\boldsymbol{x} = [x^r, x^c, x^t, x^d, x^s]$ as inputs, and fit the two processes as functions $y^m = g^m(\boldsymbol{x})$ and $y^p = g^p(\boldsymbol{x})$ separately. We use the same study area in **Figure 2**, so that $x^r$ and $x^c$ both range from 1 to 6. On each day from December 3rd to 7th, we summarize the daily order data from 7:30 to 9:30, and $x^t$ represents the index of 3-minute time intervals such that $x^t = 1$ indicates 7:30 to 7:33 and $x^t = 40$ means 9:27 to 9:30. As a result, we have $6 \times 6 \times 40 = 1{,}440$ data points for one day, and 7,200 data points in total for the five days. We consider a "five-fold" cross-validation process that takes one day's data for test and the rest four days' data for training and repeats it five times by choosing a different day for training-testing data splitting. The average results would be the inputs of performance metrics.

The matching and pickup functions $y^m = g^m(\boldsymbol{x})$ and $y^p = g^p(\boldsymbol{x})$ are fitted via three approaches in this paper for a comprehensive comparison:

- The matching learning-based approach.
- The analytical model-based approach.
- The AGPM-based approach (i.e., $g^m(\boldsymbol{x})$ and $g^p(\boldsymbol{x})$ are fitted by $f(\boldsymbol{x}) + \varepsilon$).

We list all considered models in **Table 2**. Please refer to **Appendix A** for understanding the notation and structure of different AGPMs. For matching learning-based models and AGPMs, both functions $g^m(\boldsymbol{x})$ and $g^p(\boldsymbol{x})$ are fitted; while for random perfect matching related models, we only fit $g^m(\boldsymbol{x})$ because they generally dismiss the pickup time; the Cobb-Douglas matching function is used for both $g^m(\boldsymbol{x})$ and $g^p(\boldsymbol{x})$. We choose the SE kernel for AGPMs because it provides the highest accuracy compared with other kernels listed in **Section 3.2**; readers can refer to **Appendix B** for the performance of different kernels and model structures. Note that machine learning models are estimated by python package sklearn [Pedregosa et al., 2011], deep learning models (i.e., LSTM and GNN) are coded with PyTorch [Paszke et al., 2019], the AGPMs and analytical models are developed and trained by the authors' python scripts. The models are trained and tested on a computer with a dual core-i7 processor and an 8-GB RAM. In terms of the computational efficiency, conventional machine learning models (i.e., LASSO, ARIMA, RF, XGB, SVR, and MLP) generally need 3 to 10 minutes for model training; deep learning models (LSTM, GRU, and GNN) take 4 to 5 hours to converge (on the basis of 1,000 epochs); the AGPMs need 3 to 5 hours for the likelihood maximization; and it takes 2 to 5 minutes to find the optimal parameters of analytical models (PMQ, SPMQ, and CDMF). The computational time here is for training "one-fold" dataset division within the cross-validation process.

**Table 2.** Models fitted in this paper.

| Model | Explanation |
|---|---|
| *Machine learning-based approach* | |
| LASSO | LASSO model in which alpha equals 1. |
| MLP | Multi-layer perception model, which is a basic neural network containing two hidden layers and 256 neurons in each hidden layer. |
| RF | Random forest model, and the number of trees is 100 in this paper |

| ARIMA | Autoregressive integrated moving average (ARIMA) model with structure (3, 0, 1) (Zhu et al., 2016); for each specific zone, we fit a corresponding model using its time-series data. |
|---|---|
| LSTM | Long short-term memory (LSTM) model consists of one LSTM layer with 8 features in the hidden state and two fully connected layers with activation function tanh (Zhao et al., 2017) |
| GRU | Gated recurrent unit (GRU) model composes of one GRU layer with 8 hidden features and two fully-connected layers with activation function ReLU. |
| SVR | Support vector machine model, and we use an RBF kernel in this paper |
| XGB | XGBoost model, which is an implementation of gradient boosted decision trees, and we set maximal depth as 3 and learning rate as 0.1 in this paper |
| GNN | Graph neural network model which uses the graph convolutional layer for message passing, and has a pre- and a post-processing MLP for the node features; we use 1 layer for pre-processing, 2 for message passing and 2 for post-processing (You et al., 2020) |

*Analytical approach*

| PMQ | Random perfect matching model with $g^m(x) = \min(x^{cd}, x^s)$, where $x^{cd}$ is the cumulative passenger demand that is the sum of both queued (unmatched) passenger from previous time periods and newly arrival passengers $x^d$ (Cachon et al., 2017; Yu et al., 2020) |
|---|---|
| SPMQ | Spatial PMQ model, we assume passengers can be matched with drivers from adjacent zones with $g^m(x) = \min(x^{cd}, \tilde{x}^s)$, where $\tilde{x}^s = ax^s + b\sum x_a^s$ denotes the effective supply: $x_a^s$ is the supply from adjacent zones, $a$ and $b$ are weighting factors; we treat the gap between observed and predicted values as the loss, and fit $a$ and $b$ by minimizing the loss |
| CDMF | Cobb-Douglas matching function with $g^m(x) = A_m(x^d)^{\alpha_m}(x^s)^{\beta_m}$ and $g^p(x) = A_p(x^d)^{\alpha_p}(x^s)^{\beta_p}$ (Yang et al., 2010; Zha et al., 2016); we treat the gap between observed and fitted values as the loss, and numerically estimate parameters $A_m$, $\alpha_m$, $\beta_m$, $A_p$, $\alpha_p$, and $\beta_p$ by minimizing the loss |

*AGPM based approach*

| AGPM-1 | Full GP model that takes the entire vector $x$ as the input, i.e., $f(x) = f_{SE}^{(1)}(x^{r,c,t,d,s})$ |
|---|---|
| AGPM-2 | AGPM model that considers one GP with the product kernel of spatial-temporal information and supply-demand information, i.e., $f(x) = f_{SE \times SE}^{(1)}(x^{r,c,t} \times x^{d,s})$ |
| AGPM-3 | AGPM model with four GPs of spatial, temporal, supply, and demand information, respectively, i.e., $f(x) = f_{SE}^{(1)}(x^{r,c}) + f_{SE}^{(2)}(x^t) + f_{SE}^{(3)}(x^d) + f_{SE}^{(4)}(x^s)$ |
| AGPM-4 | AGPM model that contains two GPs, one with the product kernel of spatial and supply information, the other with the product kernel of temporal and demand information, i.e., $f(x) = f_{SE \times SE}^{(1)}(x^{r,c} \times x^s) + f_{SE \times SE}^{(2)}(x^t \times x^d)$ |
| AGPM-5 | AGPM model with two GPs, both of which have a three-way product kernel: one kernel takes spatial, temporal, and demand information and the other kernel takes spatial, temporal, and supply information, i.e., $f(x) = f_{SE \times SE \times SE}^{(1)}(x^{r,c} \times x^t \times x^d) + f_{SE \times SE \times SE}^{(2)}(x^{r,c} \times x^t \times x^s)$ |

**Table 3** shows the overall accuracy of different models under the "five-fold" cross-validation approach, in which the mean absolute error (MAE), root mean squared error (RMSE), and the R-squared value ($R^2$) are utilized as the performance measurements. We use bold font to highlight models with high accuracy. Although analytical models have been generally used in ride-sourcing market analyses and led to insightful findings, the accuracy tends to be the lowest among different models. Among different machine learning models, XGB offers the highest accuracy for predicting the number of matches; GNN , LSTM, and GRU perform the best in predicting the number of pickups. It is worth noting that GNN is not so powerful in the match fitting task, indicating that GNNs could be sensitive to the data since they assume a high correlation between spatial information and the output variable. For both the matching and pickup processes, the AGPM models (especially AGPM-4 and AGPM-5) outperform most machine learning-based models and analytical models. The results demonstrate the power of AGPMs in depicting spatial-temporal ride-sourcing processes.

**Table 3** Overall accuracy of different models

| Models | Num. of Matches ($y^m$) | | | Num. of Pickups ($y^p$) | | |
|---|---|---|---|---|---|---|
| | MAE | RMSE | $R^2$ | MAE | RMSE | $R^2$ |
| LASSO | 1.829 | 2.605 | 0.828 | 2.520 | 3.483 | 0.672 |
| MLP | 1.736 | 2.481 | 0.844 | 2.329 | 3.278 | 0.709 |
| RF | 1.814 | 2.587 | 0.831 | 2.260 | 3.177 | 0.724 |
| ARIMA | 2.238 | 3.058 | 0.768 | 2.238 | 3.171 | 0.734 |
| LSTM | 1.881 | 2.632 | 0.803 | **2.195** | 3.054 | **0.735** |
| GRU | 1.763 | 2.631 | 0.835 | **2.112** | **3.005** | **0.737** |
| SVR | 1.778 | 2.593 | 0.831 | 2.383 | 3.408 | 0.687 |
| XGB | **1.706** | **2.431** | **0.850** | 2.274 | 3.166 | 0.728 |
| GNN | 2.166 | 2.964 | 0.773 | **2.149** | **3.009** | **0.736** |
| PMQ | 3.352 | 5.257 | 0.311 | - | - | - |
| SPMQ | 2.672 | 4.013 | 0.596 | - | - | - |
| CDMF | 1.961 | 2.851 | 0.794 | 2.610 | 3.678 | 0.635 |
| AGPM-1 | 1.728 | **2.441** | **0.849** | 2.289 | 3.201 | 0.723 |
| AGPM-2 | **1.701** | 2.472 | 0.847 | 2.243 | **3.131** | **0.736** |
| AGPM-3 | 1.730 | 2.455 | 0.847 | **2.232** | **3.163** | 0.729 |
| AGPM-4 | **1.676** | **2.382** | **0.858** | **2.174** | **3.069** | **0.744** |
| AGPM-5 | **1.644** | **2.347** | **0.861** | **2.058** | **2.964** | **0.740** |

"-" means no results.

Based on the results in **Table 3**, we select AGPM-5 for modeling the matching and pickup processes. According to **Section 3.2**, the parameter set of AGPM-5 is denoted as $\boldsymbol{\theta} = \left[\boldsymbol{\theta}^{(1)}, \boldsymbol{\theta}^{(2)}, \sigma_\varepsilon^2\right] = \left[\sigma_{OV}^{(1)^2}, l_{SE}^{(1),r,c}, l_{SE}^{(1),t}, l_{SE}^{(1),d}, \sigma_{OV}^{(2)^2}, l_{SE}^{(2),r,c}, l_{SE}^{(2),t}, l_{SE}^{(2),s}, \sigma_\varepsilon^2\right]$, where $\sigma_{OV}^{(d)}$ is the variance of the additive kernel (see **Appendix A** for the meaning of parameters). The values of parameters present the quantitative relationship between the number of order matches and passenger pickups and GP components with spatial, temporal, demand, and supply information. We obtain the parameter set of the matching process by taking the average parameters of the five pieces of training, which is given by $[5.4, 7.4, 20.9, 19.9, 1.6, 0.2, 41.9, 12.3, 5.1]$; in the "spatial-temporal-demand" GP (i.e., $f^{(1)}$), location, time index, and demand have a notable influence on the covariance matrix; $l_{SE}^{(2),r,c} = 0.2$ and $l_{SE}^{(2),t} = 41.9$ indicate that the "spatial-temporal-supply" GP (i.e., $f^{(2)}$) is sensitive to the location but insensitive to the time index; moreover, $\sigma_{OV}^{(1)^2} = 5.4$ and $\sigma_{OV}^{(2)^2} = 1.6$ means that the matching process (i.e., number of matches) relies more on the demand than the supply. Similarly, the parameter set of the pickup process is $[3.9, 1.0, 20.4, 29.5, 0.8, 0.2, 5.8, 1.3, 7.4]$. It implies that the number of pickups is also more sensitive to the "spatial-temporal-demand" component than the "spatial-temporal-supply" one. The results conclude that demand has a larger influence on both pickup and matching than supply in the ride-sourcing market. Comparing the matching and the pickup processes, we note that the noise variance of the latter is higher so that the pickup process has more stochasticity than the matching process.

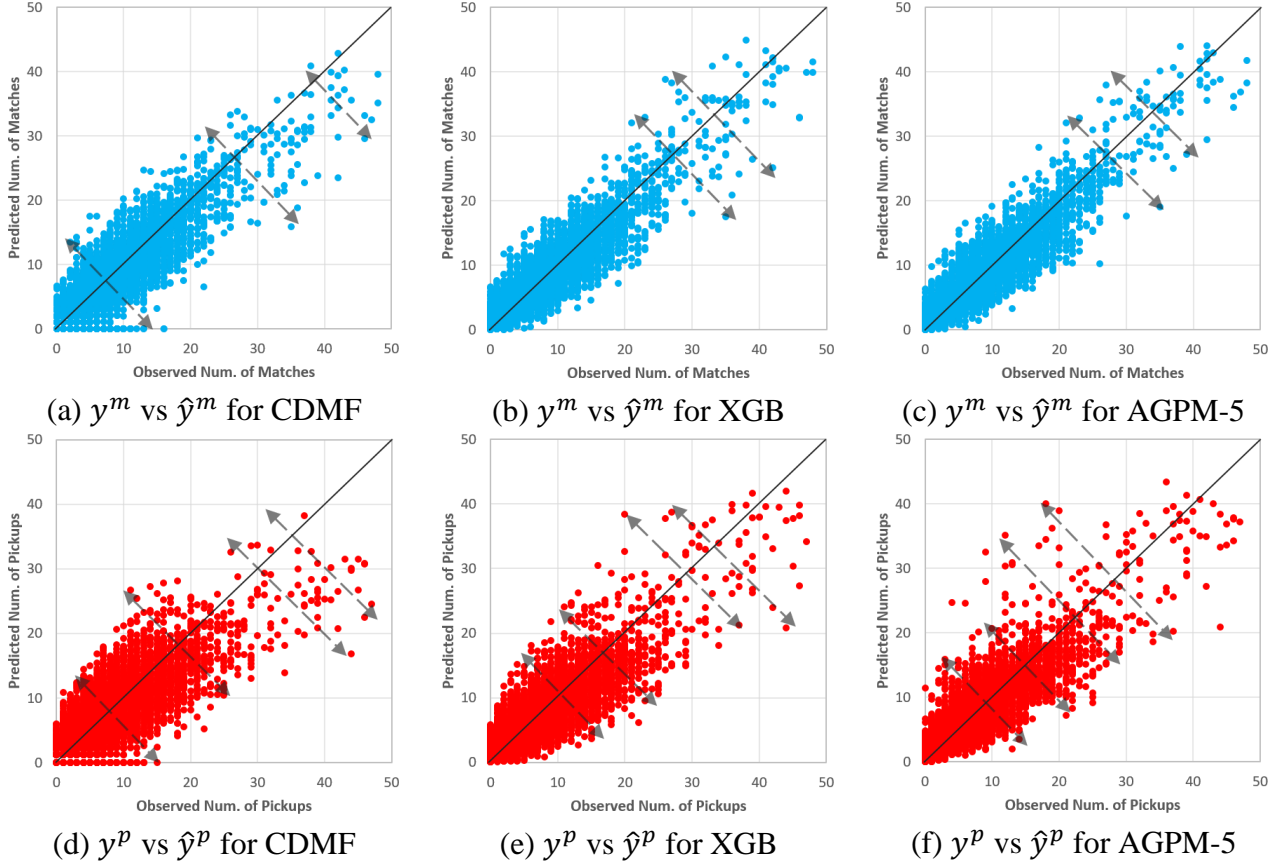

(a) $y^m$ vs $\hat{y}^m$ for CDMF  (b) $y^m$ vs $\hat{y}^m$ for XGB  (c) $y^m$ vs $\hat{y}^m$ for AGPM-5

(d) $y^p$ vs $\hat{y}^p$ for CDMF  (e) $y^p$ vs $\hat{y}^p$ for XGB  (f) $y^p$ vs $\hat{y}^p$ for AGPM-5

**Figure 4.** Observations vs predictions

Furthermore, we show the observations ($y^m$ and $y^p$) and predictions ($\hat{y}^m$ and $\hat{y}^p$) for the full dataset (from December 3rd to 7th) of models CDMF, XGB, and AGPM-5 in **Figure 4**. The CDMF models offer many "zero predictions" for $\hat{y}^m$ and $\hat{y}^p$, and tend to underestimate their values because the supply in a zone is always insufficient and even reaches zero in the morning peak. Models XGB and AGPM-5 provide trustable predictions for $y^m$; while for predicting $y^p$, there are many outliers. One can note that even though the AGPM-5 results in outliers with larger gaps, while it gives more reliable predictions when $y^p \in [0, 20]$.

## 5. Application of AGPMs in Real-World Modeling Tasks

This section shows an example of conducting modeling research tasks via AGPMs in ride-sourcing markets. We focus on the modeling of the order matching process and utilize the trained AGPM-5 ($y^m = g^m(\boldsymbol{x}) = f_{SE \times SE \times SE}^{(1)}(x^{r,c} \times x^t \times x^d) +$ $f_{SE \times SE \times SE}^{(2)}(x^{r,c} \times x^t \times x^s) + \varepsilon$) to model idle vehicle relocation strategies in the market[4]. The modeling of passenger pickup or other processes can be done in similar ways but will not be discussed in this paper.

We select December 4th as the modeling day, so that $\boldsymbol{X}^*$ is the data on December 4th, and $\boldsymbol{X}$ and $\boldsymbol{Y}$ are the data from December 3rd, 5th, 6th, and 7th. According to **Section 3.3**, the mean numbers of matches for all zones and time periods from 7:30 to 9:30 on December 4th are given by $\boldsymbol{Y}^{m*} = \boldsymbol{K}_{\boldsymbol{X}^*,\boldsymbol{X}}^F(\boldsymbol{\theta})\boldsymbol{R}_Y^F(\boldsymbol{\theta})\boldsymbol{Y}$, which is a 1,440-by-1 vector (i.e., this is the same way as we make predictions in **Section 4.3**). We refer to the estimated condition on December 4th as the baseline scenario (BS) so that $\boldsymbol{X}^*$ contains the actual spatial-temporal demand and supply information without vehicle relocations and $\boldsymbol{Y}^{m*}$ is called the BS matching results. For zone $(r, c)$, we evaluate its matching efficiency by computing the order matching queue length; the incremental matching queue during one time period is $x^{d*} - y^{m*}$ ($x^{d*}$ and $y^{m*}$ are elements of $\boldsymbol{X}^*$ and $\boldsymbol{Y}^{m*}$, respectively), and the cumulative order matching queue during the entire 40 time periods (denoted as $Q^{r,c}$) is given by

$$Q^{r,c}(\boldsymbol{X}^*, \boldsymbol{Y}^{m*}) = \sum_{t=1}^{40} q^{r,c,t} \tag{13}$$

$$q^{r,c,t} = q_0^{r,c} + \sum_{x^*|x^{r*}=r,x^{c*}=c,x^{t*}\leq t}(x^{d*} - y^{m*}) \tag{14}$$

where $q_0^{r,c}$ is the initial queue length in zone $(r, c)$ at 7:30 (obtained from the data), $q^{r,c,t}$ denotes the queue length at time period $t$. The overall cumulative queue of the market (denoted by $Q$) is computed by summing up $q^{r,c}$ across all zones:

$$Q(\boldsymbol{X}^*, \boldsymbol{Y}^{m*}) = \sum_{r=1}^{6}\sum_{c=1}^{6} Q^{r,c}(\boldsymbol{X}^*, \boldsymbol{Y}^{m*}) \tag{15}$$

The overall cumulative queue $Q$ in BS is 10,153 passengers, which might not be the most efficient situation (the zone-level cumulative queue length $Q^{r,c}$ is shown in **Figure 8(a)**). To improve matching efficiency, the platform considers implementing idle vehicle relocation strategies. Assuming the demand to be exogenous, since our matching function $y^m = g^m(\boldsymbol{x})$ takes $\boldsymbol{x} = [x^r, x^c, x^t, x^d, x^s]$ as the input, we can estimate the new matching results and summarize the performance of vehicle relocation strategy by:

- Rebalancing (revising) $x^{s*}$ of $\boldsymbol{x}^*$ in $\boldsymbol{X}^*$ while keeping the other dimensions unchanged (the revised input matrix under strategy $S$ is denoted as $\boldsymbol{X}_S^*$)
- Computing new $\boldsymbol{Y}_S^{m*}$ based on new $\boldsymbol{X}_S^*$, i.e., $\boldsymbol{Y}_S^{m*} = \boldsymbol{K}_{\boldsymbol{X}_S^*,\boldsymbol{X}}^F(\boldsymbol{\theta})\boldsymbol{R}_Y^F(\boldsymbol{\theta})\boldsymbol{Y}$.
- Computing new overall cumulative queue $Q(\boldsymbol{X}_S^*, \boldsymbol{Y}_S^{m*})$ via Eqs. (13–15).

---

[4] Note that in the modeling section, we use $y^m$ rather than $\hat{y}^m$ for the estimated value; similarly, $\boldsymbol{Y}^{m*}$ represents $\widehat{\boldsymbol{Y}}^{m*}$.

We propose three strategies: the queue-based strategy (QS), gradient-based strategy (GS), and combined strategy (CS). In each strategy, key matching efficiency metrics are computed for four time windows (i.e., 7:30 to 8:00, 8:00 to 8:30, 8:30 to 9:00, and 9:00 to 9:30); within each time window (10 time periods), we identify target zones according to the strategy and metrics, and move 10% idle vehicles from adjacent zones to these target zones.

The basic idea of QS is to relocate vehicles from the zones with a shorter matching queue to those with a longer queue. Therefore, the metric is the zone-level cumulative queue length under $X^*$ (the BS case), which is computed similarly to Eqs. (13–14) but $q^{r,c,t}$ is separately summed up for the four time windows. One zone is regarded as a target zone if its 30-minute specific cumulative queue is longer than 100 passengers. The zone-level cumulative queue computed based on $X^*$ and QS relocation strategy for the four time windows are shown in **Figure 5**, where the number in each zone denotes the cumulative queue length, target zones are marked with red fonts, and idle blue arrows denote the relocating movement of idle vehicles. For instance, as depicted in **Figure 5(a)**, for each time period from 7:30 to 8:00, we move 10% $x^s$ from zones *(1, 1)*, *(2, 1)*, and *(2, 2)* to zone *(1, 2)*.

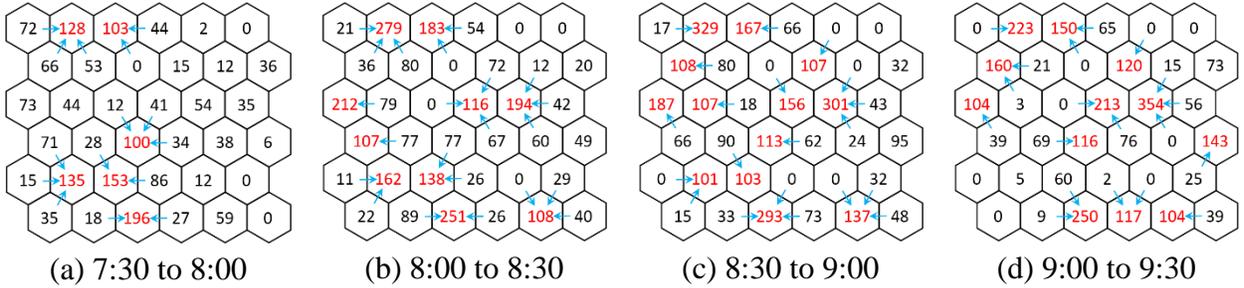

(a) 7:30 to 8:00    (b) 8:00 to 8:30    (c) 8:30 to 9:00    (d) 9:00 to 9:30

**Figure 5.** BS zone-level cumulative queue and QS vehicle movements

Since the gradient of $y^m$ with $x^s$ is heterogeneous across different time periods and zones, QS could be ineffective in cutting down the overall cumulative queue of the market. To make full use of AGPM, we compute the gradient (derivative) of $y^m$ with $x^{s*}$. Based on **Section 3.3**, the gradient of model AGPM-5 is $\frac{\mathrm{d}y^m}{\mathrm{d}x^{s*}} = W^s_{x^*,X}(\boldsymbol{\Theta}) R^F_Y(\boldsymbol{\Theta}) Y$, where $\left[W^s_{x^*,X}(\boldsymbol{\Theta})\right]_i = \frac{x^s_i - x^{s*}}{l^{(2),s^2}_{SE}} f^{(2)}(x^*, x_i|\boldsymbol{\Theta})$. The metric is the summation of $\frac{\mathrm{d}y^m}{\mathrm{d}x^{s*}}$ over different time windows for each zone under $X^*$. Once the 30-minute gradient of a zone is over 1.2, we regard it as a target zone. The zone-level gradients and the relocation strategies under GS are shown in **Figure 6**, where the number in each zone denotes the 30-minute gradient. Note that we do not move vehicles from a zone to its adjacent target zone once its gradient is between 1.0 and 1.2.

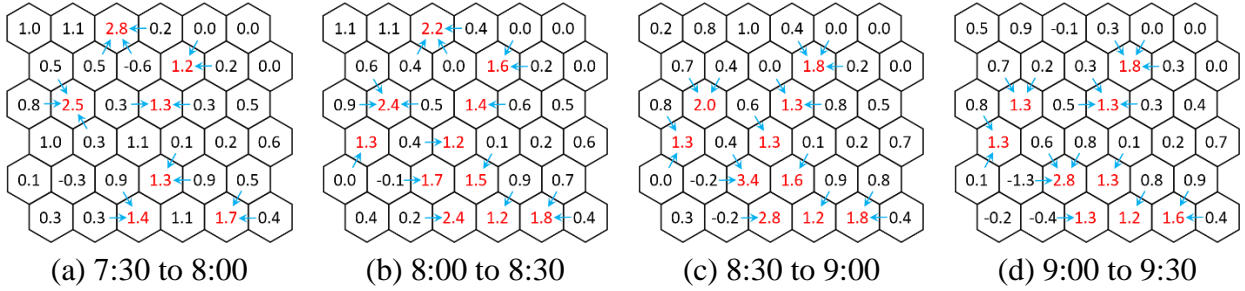

**Figure 6.** BS zone-level gradient and GS vehicle movements

GS could still be imperfect because some zones have a high queue length but low gradient, and vice versa. To jointly consider the queue length and gradient, we utilize the product of the zone-level cumulative queue length and gradient as the metric for CS. The zones with a metric over 100 are treated as targets. **Figure 7** shows the zone-level metrics and CS relocation strategy, where the numbers represent the product of the queue length and gradient.

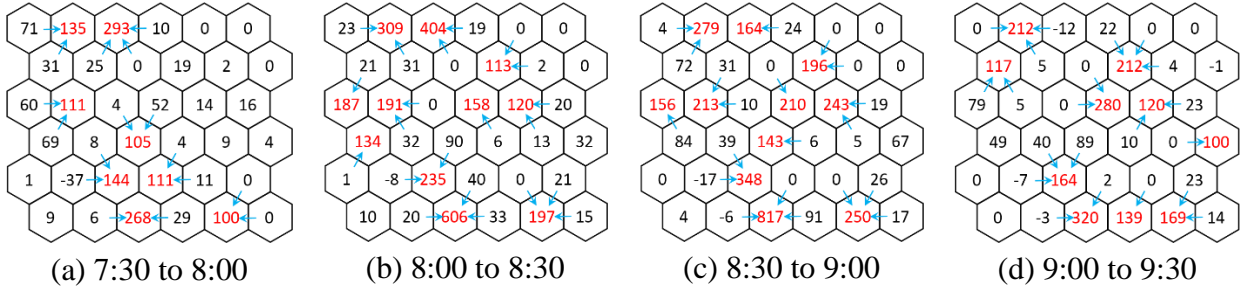

**Figure 7.** BS product of queue and gradient and CS vehicle movements

**Figure 8** presents the zone-level cumulative queue length $Q^{r,c}$ under different scenarios, i.e., $Q^{r,c}(X^*, Y^{m*})$ for BS, $Q^{r,c}(X^*_{QS}, Y^{m*}_{QS})$ for QS, $Q^{r,c}(X^*_{GS}, Y^{m*}_{GS})$ for GS, and $Q^{r,c}(X^*_{CS}, Y^{m*}_{CS})$ for CS. We mark the number with red/blue font if the queue length increases/decreases over 10% compared with the BS. Overall, we have $Q(X^*, Y^{m*}) = 10,153$, $Q(X^*_{QS}, Y^{m*}_{QS}) = 9,741$, $Q(X^*_{QS}, Y^{m*}_{QS}) = 9,065$, and $Q(X^*_{CS}, Y^{m*}_{CS}) = 9,258$ passengers; QS can reduce the total cumulative queue by around 3%, GS has the best performance by decreasing the total cumulative queue by 10%, while CS takes an intermediate effect that reduces the queue by 7%. Although the impact of GS is significant, it might ignore the balance of service qualities among zones; for instance, passengers in zone *(1, 2)* still suffer from serious queueing. This drawback is notably addressed in CS (comparing **Figure 8(c)** with **Figure 8(d)**).

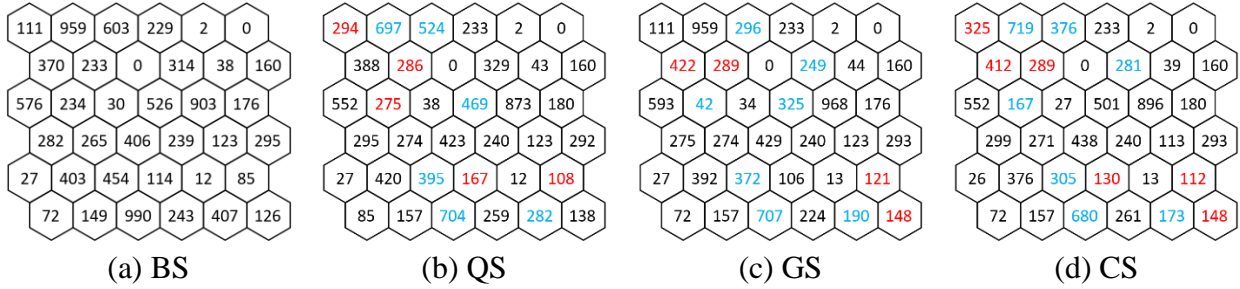

**Figure 8.** Zone-level cumulative queue length under different strategies

To this end, we have illustrated the application of AGPM in operational analyses in the ride-sourcing market. As an analytically tractable model, one can make good use of the formulas of mean, variance, and their derivatives in many research tasks. Note that the results in **Figure 8** could be different from ground truth data once the same relocation strategies are truly adopted in the real world. However, AGPM provides a data-driven approach for conducting modeling tasks. To obtain more solid results and tackle more comprehensive research problems, we can involve additional data and use multi-output AGPMs that incorporate passenger demand or driver supply as endogenous variables. Moreover, we can integrate AGPMs with advanced optimization methods such as reinforcement learning (Zhu et al., 2021a).

## 6. Conclusions

This paper proposes a novel data-driven approach for fitting and modeling spatial-temporal stochastic ride-sourcing processes via additive Gaussian Process models (AGPMs). We analyze and models the order matching and passenger pickup processes in ride-sourcing markets, which reflect the core features of ride-sourcing services. Based on ride-sourcing order data collected in Hangzhou, China, we find that, due to spatial-temporal imbalance between passenger demand and driver supply, the matching and pickup processes are too complicated to be fully characterized by general queueing or analytical models. We fit the processes by treating spatial-temporal information and demand-supply information as the exogenous (input) covariates and the number of matches and pickups as the endogenous variables (outputs). We train machine learning models, analytical models, and AGPMs for comprehensive comparisons. The results indicate that the AGPMs are superior to other widely-used machine learning models and analytical models in offering accurate estimations for the number of matches and pickups. Furthermore, the trained AGPM is utilized to model idle vehicle relocation strategies for improving the order matching efficiency in the market, which has demonstrated the capability of the data-driven approach in modeling tasks such as market analysis and design of operational strategies.

The major contributions of this study include 1) we propose a data-driven model, i.e., the AGPM, for modeling ride-sourcing markets and the AGPM is superior in both offering high fitting/prediction accuracy and providing analytically tractable expressions; 2) we show the fitting and modeling power of the AGPM via empirical studies based on real-world data; 3) we link analytical models with real-world ride-sourcing data and find that they might not be suitable for depicting the actual matching and pickup processes; while previous studies generally utilized analytical models in ride-sourcing studies without testifying the goodness of fit for data; 4) we test the performance of widely-used machine learning methods in predicting the number of matches and pickups, which has widened the application of machine learning in ride-sourcing studies.

There are some limitations in this paper. First, the number of matches and the number of pickups are modeled separately. With advanced methods such as multi-task GPs, we will attempt to model the two processes in the future jointly. Second, the Hangzhou DiDi order data has not included order cancelations, which could dismiss the impact of cancel behaviors on the matching process. This issue shall be addressed once the dataset is fulfilled. Third, since the AGPM based on conventional GPs is computationally inefficient, the model requires a large computer RAM for applications in large-scale networks. We will use large-scale oriented methods, such as sparse approximation and approximated AGPM to extend our work on larger study areas. The capability of modeling matching and pickup processes in large areas has promising applications: 1) to predict large-scale demand-supply imbalance for monitoring purposes; 2) to be integrated with reinforcement learning approaches for analyzing and designing dynamic operational strategies such as surge pricing and subsidy. Moreover, we will apply the AGPM and advanced GP-related models to fit and model other spatial-temporal processes in transportation systems, such as the rent and return process between bikes and riders in bike-sharing systems (Xu et al., 2022) and the interactive relationship between autonomous vehicles and human-driven vehicles in mixed traffic flow scenarios (Lu et al., 2021); although the observed data of the latter case is limited, the AGPM can be adopted to approximate simulated mixed flow dynamics.

## Acknowledgments

The work described in this paper was partially supported by Hong Kong Research Grants Council under project 16208920, Alibaba-Zhejiang University Joint Research Institute of Frontier Technologies, and Hong Kong University of Science and Technology– DiDi Chuxing (HKUST-DiDi) Joint Laboratory.

## References

Agatz, N., Erera, A. L., Savelsbergh, M. W., & Wang, X. (2011). Dynamic ride-sharing: A simulation study in metro Atlanta. *Procedia-Social and Behavioral Sciences*, *17*, 532–550.

Allahviranloo, M., & Recker, W. (2013). Daily activity pattern recognition by using support vector machines with multiple classes. *Transportation Research Part B: Methodological*, *58*, 16–43.

Bai, J., So, K. C., Tang, C. S., Chen, X. (Michael), & Wang, H. (2019). Coordinating Supply and Demand on an On-Demand Service Platform with Impatient Customers. *Manufacturing & Service Operations Management*, *21*(3), 556–570.

Baldacci, R., Maniezzo, V., & Mingozzi, A. (2004). An exact method for the car pooling problem based on lagrangean column generation. *Operations Research*, *52*(3), 422–439.

Banerjee, S., Johari, R., & Riquelme, C. (2016). Dynamic pricing in ridesharing platforms. *ACM SIGecom Exchanges*, *15*(1), 65–70.

Battifarano, M., & Qian, Z. S. (2019). Predicting real-time surge pricing of ride-sourcing companies. *Transportation Research Part C: Emerging Technologies*, *107*, 444–462.

Cachon, G. P., Daniels, K. M., & Lobel, R. (2017). The Role of Surge Pricing on a Service Platform with Self-Scheduling Capacity. *Manufacturing & Service Operations Management, 19*(3), 368–384.

Cantarella, G. E., & de Luca, S. (2005). Multilayer feedforward networks for transportation mode choice analysis: An analysis and a comparison with random utility models. *Transportation Research Part C: Emerging Technologies*, *13*(2), 121–155.

Castillo, J. C., Knoepfle, D., & Weyl, G. (2017). Surge Pricing Solves the Wild Goose Chase. *Proceedings of the 2017 ACM Conference on Economics and Computation*, 241–242.

Chen, P. W., & Nie, Y. M. (2017). Connecting e-hailing to mass transit platform: Analysis of relative spatial position. *Transportation Research Part C: Emerging Technologies*, *77*, 444-461.

Cheng, L., Ramchandran, S., Vatanen, T., Lietzén, N., Lahesmaa, R., Vehtari, A., & Lähdesmäki, H. (2019). An additive Gaussian process regression model for interpretable non-parametric analysis of longitudinal data. *Nature Communications*, *10*(1), 1–11.

Feng, G., Kong, G., & Wang, Z. (2021). We are on the way: Analysis of on-demand ride-hailing systems. *Manufacturing & Service Operations Management*, *23*(5), 1237–1256.

Ganji, A., Shekarrizfard, M., Harpalani, A., Coleman, J., & Hatzopoulou, M. (2020). Methodology for spatio-temporal predictions of traffic counts across an urban road network and generation of an on-road greenhouse gas emission inventory. *Computer-Aided Civil and Infrastructure Engineering*, *35*(10), 1063–1084.

Idé, T., & Kato, S. (2009). Travel-time prediction using Gaussian process regression: A trajectory-based approach. *Proceedings of the 2009 SIAM International Conference on Data Mining*, 1185–1196.

Holler, J., Vuorio, R., Qin, Z., Tang, X., Jiao, Y., Jin, T., ... & Ye, J. (2019). Deep reinforcement learning for multi-driver vehicle dispatching and repositioning problem. *2019 IEEE International Conference on Data Mining (ICDM)*, 1090–1095.

Jung, J., Jayakrishnan, R., & Park, J. Y. (2016). Dynamic shared-taxi dispatch algorithm with hybrid-simulated annealing. *Computer-Aided Civil and Infrastructure Engineering*, *31*(4), 275-291.

Karlaftis, M. G., & Vlahogianni, E. I. (2011). Statistical methods versus neural networks in transportation research: Differences, similarities and some insights. *Transportation Research Part C: Emerging Technologies*, *19*(3), 387–399.

Ke, J., Li, X., Yang, H., & Yin, Y. (2021). Pareto-efficient solutions and regulations of congested ride-sourcing markets with heterogeneous demand and supply. *Transportation Research Part E: Logistics and Transportation Review*, *154*, 102483.

Ke, J., Yang, H., Li, X., Wang, H., & Ye, J. (2020). Pricing and equilibrium in on-demand ride-pooling markets. *Transportation Research Part B: Methodological*, *139*, 411–431.

Kupilik, M., & Witmer, F. (2018). Spatio-temporal violent event prediction using Gaussian process regression. *Journal of Computational Social Science*, *1*(2), 437–451.

Lin, H., Li, L., & Wang, H. (2020). Survey on research and application of support vector machines in intelligent transportation system. *Journal of Frontiers of Computer Science and Technology*, *14*(6), 901–917.

Lu, L., Zhu, Z., Guo, P., & He, Q. C. (2021). Service Operations for Mixed Autonomous Paradigm: Lane Design and Subsidy. *Production and Operations Management*, https://doi.org/10.1111/poms.13633

Ma, P., Konomi, B. A., & Kang, E. L. (2019). An additive approximate Gaussian process model for large spatio-temporal data. *Environmetrics*, *30*(8), e2569.

Masoud, N., & Jayakrishnan, R. (2017). A decomposition algorithm to solve the multi-hop peer-to-peer ride-matching problem. *Transportation Research Part B: Methodological*, *99*, 1–29.

Neumann, M., Kersting, K., Xu, Z., & Schulz, D. (2009). Stacked Gaussian process learning. *2009 Ninth IEEE International Conference on Data Mining*, 387–396.

Nourinejad, M., & Ramezani, M. (2020). Ride-Sourcing modeling and pricing in non-equilibrium two-sided markets. *Transportation Research Part B: Methodological*, *132*, 340–357.

Paszke, A., Gross, S., Massa, F., Lerer, A., Bradbury, J., Chanan, G., ... & Chintala, S. (2019). Pytorch: An imperative style, high-performance deep learning library. *Advances in neural information processing systems*, *32*, 8026-8037.

Pedregosa, F., Varoquaux, G., Gramfort, A., Michel, V., Thirion, B., Grisel, O., ... & Duchesnay, E. (2011). Scikit-learn: Machine learning in Python. *Journal of Machine Learning research*, *12*, 2825–2830.

Peng, H., Wang, H., Du, B., Bhuiyan, M. Z. A., Ma, H., Liu, J., ... & Philip, S. Y. (2020). Spatial temporal incidence dynamic graph neural networks for traffic flow forecasting. *Information Sciences*, *521*, 277–290.

Ramezani, M., & Nourinejad, M. (2018). Dynamic modeling and control of taxi services in large-scale urban networks: A macroscopic approach. *Transportation Research Part C: Emerging Technologies*, *94*, 203-219.

Rodrigues, F., & Pereira, F. C. (2018). Heteroscedastic Gaussian processes for uncertainty modeling in large-scale crowdsourced traffic data. *Transportation Research Part C: Emerging Technologies*, *95*, 636–651.

Saadi, I., Wong, M., Farooq, B., Teller, J., & Cools, M. (2017). An investigation into machine learning approaches for forecasting spatio-temporal demand in ride-hailing service. *ArXiv:1703.02433*.

Senanayake, R., O'Callaghan, S., & Ramos, F. (2016). Predicting spatio-temporal propagation of seasonal influenza using variational Gaussian process regression. In *Proceedings of the AAAI Conference on Artificial Intelligence* (Vol. 30, No. 1).

Sheibani, M., & Ou, G. (2021). The development of Gaussian process regression for effective regional post-earthquake building damage inference. *Computer-Aided Civil and Infrastructure Engineering*, *36*(3), 264–288.

Tafreshian, A., Masoud, N., & Yin, Y. (2020). Frontiers in service science: Ride matching for peer-to-peer ride sharing: A review and future directions. *Service Science*, *12*(2-3), 44–60.

Taylor, T. A. (2018). On-Demand Service Platforms. *Manufacturing & Service Operations Management*, *20*(4), 704–720.

Tseng, F.-H., Hsueh, J.-H., Tseng, C.-W., Yang, Y.-T., Chao, H.-C., & Chou, L.-D. (2018). Congestion Prediction with Big Data for Real-Time Highway Traffic. *IEEE Access*, *6*, 57311–57323.

Venkitaraman, A., Chatterjee, S., & Handel, P. (2020). Gaussian processes over graphs. In *ICASSP 2020-2020 IEEE International Conference on Acoustics, Speech and Signal Processing (ICASSP)*, pp. 5640–5644.

Wang, H., & Yang, H. (2019). Ridesourcing systems: A framework and review. *Transportation Research Part B: Methodological*, *129*, 122–155.

Wang, X., Liu, W., Yang, H., Wang, D., & Ye, J. (2020). Customer behavioural modelling of order cancellation in coupled ride-sourcing and taxi markets. *Transportation Research Part B: Methodological*, *132*, 358–378.

Wang, X., Ma, Y., Wang, Y., Jin, W., Wang, X., Tang, J., ... & Yu, J. (2020). Traffic flow prediction via spatial temporal graph neural network. In *Proceedings of The Web Conference 2020*, pp. 1082–1092.

Wang, Y., Zhang, D., Liu, Y., Dai, B., & Lee, L. H. (2019). Enhancing transportation systems via deep learning: A survey. *Transportation Research Part C: Emerging Technologies*, *99*, 144–163.

Wei, B., Saberi, M., Zhang, F., Liu, W., & Waller, S. T. (2020). Modeling and managing ridesharing in a multi-modal network with an aggregate traffic representation: a doubly dynamical approach. *Transportation Research Part C: Emerging Technologies*, *117*, 102670.

Williams, C. (2006). *Gaussian processes for machine learning*. Taylor & Francis Group.

Wu, C.-H., Ho, J.-M., & Lee, D. T. (2004). Travel-Time Prediction with Support Vector Regression. *IEEE Transactions on Intelligent Transportation Systems*, *5*(4), 276–281.

Xie, Y., Zhao, K., Sun, Y., & Chen, D. (2010). Gaussian processes for short-term traffic volume forecasting. *Transportation Research Record*, *2165*(1), 69–78.

Xu, K., Saberi, M., & Liu, W. (2022). Dynamic pricing and penalty strategies in a coupled market with ridesourcing service and taxi considering time-dependent order cancellation behaviour. *Transportation Research Part C: Emerging Technologies*, *138*, 103621.

Xu, M., Di, Y., Zhu, Z., Yang, H., & Chen, X. (2022). Designing van-based mobile battery swapping and rebalancing services for dockless ebike-sharing systems based on the dueling double deep Q-network. *Transportation Research Part C: Emerging Technologies*, *138*, 103620.

Xu, Z., Yin, Y., & Ye, J. (2020). On the supply curve of ride-hailing systems. *Transportation Research Part B: Methodological*, *132*, 29–43.

Xu, Z., Yin, Y., & Zha, L. (2017). Optimal parking provision for ride-sourcing services. *Transportation Research Part B: Methodological*, *105*, 559–578.

Yang, H., Leung, C. W. Y., Wong, S. C., & Bell, M. G. H. (2010). Equilibria of bilateral taxi–customer searching and meeting on networks. *Transportation Research Part B: Methodological*, *44*(8–9), 1067–1083.

Yang, H., & Yang, T. (2011). Equilibrium properties of taxi markets with search frictions. *Transportation Research Part B: Methodological*, *45*(4), 696–713.

Yao, B., Chen, C., Cao, Q., Jin, L., Zhang, M., Zhu, H., & Yu, B. (2017). Short-term traffic speed prediction for an urban corridor. *Computer-Aided Civil and Infrastructure Engineering*, *32*(2), 154-169.

Yao, H., Wu, F., Ke, J., Tang, X., Jia, Y., Lu, S., Gong, P., Ye, J., & Li, Z. (2018). Deep Multi-View Spatial-Temporal Network for Taxi Demand Prediction. *ArXiv:1802.08714*

Ye, J., Zhao, J., Ye, K., & Xu, C. (2020). How to Build a Graph-Based Deep Learning Architecture in Traffic Domain: A Survey. *IEEE Transactions on Intelligent Transportation Systems*, 1–21.

You, J., Ying, Z., & Leskovec, J. (2020). Design space for graph neural networks. *Advances in Neural Information Processing Systems*, *33*.

Yu, J. J., Tang, C. S., Max Shen, Z.-J., & Chen, X. M. (2020). A Balancing Act of Regulating On-Demand Ride Services. *Management Science*, *66*(7), 2975–2992.

Yuan, Y., Zhang, Z., Yang, X. T., & Zhe, S. (2021). Macroscopic traffic flow modeling with physics regularized Gaussian process: A new insight into machine learning applications in transportation. *Transportation Research Part B: Methodological*, *146*, 88–110.

Zeng, X., & Zhang, Y. (2013). Development of Recurrent Neural Network Considering Temporal-Spatial Input Dynamics for Freeway Travel Time Modeling: Temporal-spatial travel time modeling. *Computer-Aided Civil and Infrastructure Engineering*, *28*(5), 359–371.

Zha, L., Yin, Y., & Yang, H. (2016). Economic analysis of ride-sourcing markets. *Transportation Research Part C: Emerging Technologies*, *71*, 249–266.

Zhang, W., Honnappa, H., & Ukkusuri, S. V. (2020). Modeling urban taxi services with e-hailings: A queueing network approach. *Transportation Research Part C: Emerging Technologies*, *113*, 332–349.

Zhang, Y., & Xie, Y. (2007). Forecasting of Short-Term Freeway Volume with v-Support Vector Machines. *Transportation Research Record: Journal of the Transportation Research Board*, *2024*(1), 92–99.

Zhao, Z., Chen, W., Wu, X., Chen, P. C., & Liu, J. (2017). LSTM network: a deep learning approach for short-term traffic forecast. *IET Intelligent Transport Systems*, *11*(2), 68-75.

Zhou, X., Yu, W., & Sullivan, W. C. (2016). Making pervasive sensing possible: Effective travel mode sensing based on smartphones. *Computers, Environment and Urban Systems*, *58*, 52–59.

Zhu, Z., Ke, J., & Wang, H. (2021a). A mean-field Markov decision process model for spatial-temporal subsidies in ride-sourcing markets. *Transportation Research Part B: Methodological*, *150*, 540-565.

Zhu, Z., Peng, B., Xiong, C., & Zhang, L. (2016). Short-term traffic flow prediction with linear conditional Gaussian Bayesian network. *Journal of Advanced Transportation*, *50*(6), 1111-1123.

Zhu, Z., Qin, X., Ke, J., Zheng, Z., & Yang, H. (2020). Analysis of multi-modal commute behavior with feeding and competing ridesplitting services. *Transportation Research Part A: Policy and Practice*, *132*, 713–727.

Zhu, Z., Sun, L., Chen, X., & Yang, H. (2021b). Integrating probabilistic tensor factorization with Bayesian supervised learning for dynamic ridesharing pattern analysis. *Transportation Research Part C: Emerging Technologies*, *124*, 102916.

Zhu, Z., Xu, A., He, Q. C., & Yang, H. (2021c). Competition between the transportation network company and the government with subsidies to public transit riders. *Transportation Research Part E: Logistics and Transportation Review*, *152*, 102426.

## Appendix A. Example Kernels and Notations of AGPM Structures

The following kernels are usually used in developing GP-based models for spatial-temporal research tasks (Williams, 2006; Cheng et al., 2019):

- The squared exponential (SE) kernel, which is generally used in continuous features and spatial coordinates:

$$k_{SE}(\boldsymbol{x}, \boldsymbol{x}'|\boldsymbol{\theta}_{SE}) = \sigma_{SE}^2 \exp\left(-\frac{\|\boldsymbol{x} - \boldsymbol{x}'\|^2}{2l_{SE}^2}\right)$$

where $\sigma_{SE}^2$ is the variance and $l_{SE} > 0$ is the length-scale parameter, so that $\boldsymbol{\theta}_{SE} = [\sigma_{SE}^2, l_{SE}]$.

- The periodic (PE) kernel that can be used for continuous covariates such as timestamps:

$$k_{PE}(\boldsymbol{x}, \boldsymbol{x}'|\boldsymbol{\theta}_{PE}) = \sigma_{PE}^2 \exp\left(-\frac{2\sin^2\big((\boldsymbol{x} - \boldsymbol{x}')/\gamma_{PE}\big)}{l_{PE}^2}\right)$$

where $\sigma_{PE}^2$ is the variance, $l_{PE} > 0$ is the square of length-scale parameter, and $\gamma_{PE}$ is the periodic length; therefore, $\boldsymbol{\theta}_{PE} = [\sigma_{PE}^2, l_{PE}, \gamma_{PE}]$.

- The categorical (CA) kernel that can be used for categorical covariates:

$$k_{CA}(\boldsymbol{x}, \boldsymbol{x}'|\boldsymbol{\theta}_{CA}) = \begin{cases} \sigma_{CA}^2 & if \ \boldsymbol{x} = \boldsymbol{x}' \\ 0 & otherwise \end{cases}$$

where $\boldsymbol{\theta}_{CA} = [\ \sigma_{CA}^2]$ only contains the variance.

- The Ornstein-Uhlenbeck (OU) kernel that works for continuous covariates:

$$k_{OU}(\boldsymbol{x}, \boldsymbol{x}'|\boldsymbol{\theta}_{OU}) = \sigma_{OU}^2 \exp\left(-\frac{\|\boldsymbol{x} - \boldsymbol{x}'\|}{l_{OU}}\right)$$

where $\sigma_{OU}^2$ is the variance and $l_{OU} > 0$ is the length-scale parameter, so that $\boldsymbol{\theta}_{OU} = [\sigma_{OU}^2, l_{OU}]$.

* The binary (BI) kernel that is adopted for binary covariates:

$$k_{BI}(\boldsymbol{x}, \boldsymbol{x}'|\boldsymbol{\theta}_{BI}) = \begin{cases} \sigma_{BI}^2 & if\ \boldsymbol{x} = \boldsymbol{1}\ and\ \boldsymbol{x}' = \boldsymbol{1} \\ 0 & otherwise \end{cases}$$

where $\boldsymbol{\theta}_{BI} = [\ \sigma_{BI}^2]$ only contains the variance.

We use the subscript of a GP to present its kernel; for instance, $f_{SE}^{(d)}(\boldsymbol{x}^{(d)})$ means the kernel of $f^{(d)}$ is an SE. In addition to using a single type of kernel, we can also use the production of two or more kernels for a specific GP in the AGPM. The production kernel has one overall-shared variance, and each factor kernel has its other parameters (e.g., length-scale). For instance, let $\boldsymbol{x}^{(d)} = [x^u, x^v, x^w]$, that is, the $d$th GP is computed based on dimensions $u$, $v$, and $w$ of input vector $\boldsymbol{x}$, the kernel of the GP with notation $f_{SE \times OU}^{(d)}(x^{u,v} \times x^w)$ is given by $\sigma_{OV}^{(d)2} \exp\left(-\frac{\|[x^u, x^v] - [x^u, x^v]'\|^2}{2l_{SE}^{(d),u,v2}}\right) \exp\left(-\frac{\|x^w - x^{w'}\|}{l_{OU}^{(d),w}}\right)$, where $\sigma_{OV}^{(d)2}$ represents the overall shared variance for GP $f_{SE}^{(d)}(\boldsymbol{x}^{(d)})$, $l_{SE}^{(d),u,v}$ and $l_{OU}^{(d),w}$ denote the length-scale related parameters for the SE kernel and OU kernel, respectively.

## Appendix B. Extra Results with Different Types of Kernels

We provide the results of AGPMs with different kernels in this appendix. Kernels SE, OU, and PE are utilized because they are more suitable for continuous variables. The overall accuracy under the "five-fold" cross-validation is shown in **Table A.1**, where the models with SE kernels are used in the case study of this paper.

**Table A.1.** Overall Accuracy of AGPMs with Different Kernels

| Models | Kernels | Num. of Matches ($y^m$) | | | Num. of Pickups ($y^p$) | | |
|---|---|---|---|---|---|---|---|
| | | MAE | RMSE | $R^2$ | MAE | RMSE | $R^2$ |
| AGPM-1 | $f_{SE}^{(1)}(x^{r,c,t,d,s})$ | 1.728 | 2.441 | 0.849 | 2.289 | 3.201 | 0.723 |
| AGPM-2 | $f_{SE \times SE}^{(1)}(x^{r,c,t} \times x^{d,s})$ | 1.701 | 2.472 | 0.847 | 2.243 | 3.131 | 0.736 |
| AGPM-3 | $f_{SE}^{(1)}(x^{r,c}) + f_{SE}^{(2)}(x^t)$ $+ f_{SE}^{(3)}(x^d) + f_{SE}^{(4)}(x^s)$ | 1.730 | 2.455 | 0.847 | 2.232 | 3.163 | 0.729 |
| AGPM-4 | $f_{SE \times SE}^{(1)}(x^{r,c} \times x^s)$ $+ f_{SE \times SE}^{(2)}(x^t \times x^d)$ | 1.676 | 2.382 | 0.858 | 2.174 | 3.069 | 0.744 |
| AGPM-5 | $f_{SE \times SE \times SE}^{(1)}(x^{r,c} \times x^t \times x^d)$ $+ f_{SE \times SE \times SE}^{(2)}(x^{r,c} \times x^t \times x^s)$ | 1.644 | 2.347 | 0.861 | 2.058 | 2.964 | 0.740 |
| AGPM-1 | $f_{OU}^{(1)}(x^{r,c,t,d,s})$ | 1.741 | 2.488 | 0.842 | 2.325 | 3.276 | 0.710 |
| AGPM-2 | $f_{OU \times OU}^{(1)}(x^{r,c,t} \times x^{d,s})$ | 1.732 | 2.466 | 0.846 | 2.254 | 3.181 | 0.724 |
| AGPM-3 | $f_{OU}^{(1)}(x^{r,c}) + f_{OU}^{(2)}(x^t)$ $+ f_{OU}^{(3)}(x^d) + f_{OU}^{(4)}(x^s)$ | 1.734 | 2.470 | 0.847 | 2.241 | 3.182 | 0.725 |
| AGPM-4 | $f_{OU \times OU}^{(1)}(x^{r,c} \times x^s)$ $+ f_{OU \times OU}^{(2)}(x^t \times x^d)$ | 1.679 | 2.402 | 0.853 | 2.192 | 3.105 | 0.736 |
| AGPM-5 | $f_{OU \times OU \times OU}^{(1)}(x^{r,c} \times x^t \times x^d)$ $+ f_{OU \times OU \times OU}^{(2)}(x^{r,c} \times x^t \times x^s)$ | 1.655 | 2.375 | 0.855 | 2.123 | 3.094 | 0.733 |
| AGPM-3 | $f_{SE}^{(1)}(x^{r,c}) + f_{PE}^{(2)}(x^t)$ $+ f_{SE}^{(3)}(x^d) + f_{SE}^{(4)}(x^s)$ | 1.775 | 2.477 | 0.844 | 2.280 | 3.224 | 0.719 |
| AGPM-4 | $f_{SE \times SE}^{(1)}(x^{r,c} \times x^s)$ $+ f_{PE \times SE}^{(2)}(x^t \times x^d)$ | 1.705 | 2.429 | 0.848 | 2.233 | 3.152 | 0.731 |
| AGPM-5 | $f_{SE \times PE \times SE}^{(1)}(x^{r,c} \times x^t \times x^d)$ $+ f_{SE \times PE \times SE}^{(2)}(x^{r,c} \times x^t \times x^s)$ | 1.681 | 2.399 | 0.854 | 2.151 | 3.121 | 0.734 |